\documentclass[preprint,11pt]{aastex}
\usepackage{amssymb,amsmath}
\usepackage{color,nicefrac}
\def\pg{PG~1116+215}
\def\hst{\it HST\rm}
\def\fuse{\it FUSE\rm}
\def\nodata{---}

\def\ovi{O~VI}
\def\ovii{O~VII}
\def\oviii{O~VIII}

\def\hi{H~I}
\def\nvii{N~VII}
\def\chandra{\it Chandra\rm}
\def\xmm{\it XMM-Newton\rm}
\begin{document}
\title{A possible Chandra and Hubble Space Telescope detection of extragalactic  WHIM towards \pg}

\author{M. Bonamente\altaffilmark{1,2}, J. Nevalainen\altaffilmark{3}, 
E. Tilton\altaffilmark{4}, J. Liivam\"agi\altaffilmark{3},
E. Tempel\altaffilmark{3}, P. Hein\"am\"aki\altaffilmark{5} and T. Fang\altaffilmark{6}
}

\altaffiltext{1}{Department of Physics, University of Alabama in Huntsville,
Huntsville, Al}
\altaffiltext{2}{NASA National Space Science and Technology Center, Huntsville, Al}
\altaffiltext{3}{Tartu Observatory, Observatooriumi 1, 61602 T{\~{o}}ravere, Estonia}
\altaffiltext{4}{CASA, Department of Astrophysical \& Planetary Sciences, University of Colorado, Boulder, CO 80309}
\altaffiltext{5}{Tuorla Observatory, V\"ais\"al\"antie 20, 21500,
Piikki\"o, Finland}
\altaffiltext{6}{Department of Astronomy and Institute for Theoretical Physics and Astrophysics, Xiamen University, Xiamen, Fujian, China }

\begin{abstract}
We have analyzed \chandra\ LETG and \xmm\ RGS spectra towards the $z=0.177$ quasar \pg, 
a sightline that
is rendered particularly interesting by the \hst\ detection of several
\ovi\ and \hi\ broad Lyman-$\alpha$ absorption lines that may be associated
with the warm--hot intergalactic medium.

We performed a search for resonance K$\alpha$ absorption lines from \ovii\ and \oviii\ at the 
redshifts of the detected far--ultraviolet lines.
We detected an absorption line in the Chandra spectra at 5.2 $\sigma$ 
confidence level at wavelengths corresponding to \oviii\ K$\alpha$ 
at $z = 0.0911\pm0.0004\pm0.0005$ (statistical followed by systematic error). 
This redshift is within 3 $\sigma$ of that of  a \hi\ 
broad Lyman-$\alpha$ of $b \simeq 130$~km/s (corresponding to a temperature of
$\log T(K) \simeq 6.1$) at $z = 0.09279\pm0.00005$.
We have also analyzed the available \xmm\ RGS data towards \pg. Unfortunately, 
the \xmm\ data are not suitable to investigate this line because of instrumental
features at the wavelenghts of interest.
At the same redshift, the \chandra\ and \xmm\ spectra
have \ovii\ K$\alpha$ absorption line features of significance
1.5 $\sigma$ and 1.8 $\sigma$, respectively.

We also analyzed the available SDSS spectroscopic galaxy survey
data towards \pg\ in the redshift range
of interest. We found evidence for a galaxy filament that intersect
the \pg\ sightline and additional galaxy structures that may 
host WHIM. The \hi\ BLA and the \oviii\ K$\alpha$ absorbers are within a
few  Mpc of the filament (assuming that redshifts track Hubble flow distances) 
or consistent with gas accreting onto the filament from either direction 
relative to the sightline with velocities of a few $\times 100$~km/s.

The combination of \hst, \chandra, \xmm\ and SDSS data
indicates that we have likely detected a multi--temperature WHIM at $z\simeq 0.091-0.093$
towards \pg.
The  \oviii\ K$\alpha$ absorption line indicates gas at high temperature, 
$\log T(K) \geq 6.4$,
with a total column density of order $\log N_H(\text{cm}^2) \geq 20$
and a baryon overdensity $\delta_b \sim 100-1000$ for sightline lengths of
$L=1-10$~Mpc. This detection
highlights the importance of BLA absorption lines as possible signposts
of high--temperature WHIM filaments.

\end{abstract}

\section{Introduction: the search for missing baryons at low redshift}

The intergalactic medium (IGM) contains the vast majority of the universe's baryons
at all redshifts \citep[][and references therein]{shull2012}. At high redshift, the bulk
 of this mass is in the photoionized phase that gives rise to the Lyman-$\alpha$ forest, but 
at lower redshifts a diffuse warm--hot intergalactic medium (WHIM) at temperatures
 $\log T(K)=5-7$ is predicted to contain approximately 50\% of 
the baryons in the universe \citep[e.g.][]{dave2001,cen1999}.
Absorption--line spectroscopy in the far ultraviolet (FUV) has proven a successful
means for studying the IGM at low redshift, and a number of surveys have used 
space--based observations to place constraints on the baryonic content of its 
different phases \citep[e.g.,][]{danforth2008,tripp2008,tilton2012,shull2012}. 
Nonetheless, these surveys have detected only a fraction of the expected baryons.
The detection of WHIM gas remains particularly incomplete, owing to the difficulty 
associated with detecting the broad Lyman-$\alpha$ H~I absorption (BLA) lines 
and highly ionized metal lines (such as K$\alpha$ from \ovii\ and \oviii)
that are characteristic of diffuse gas at WHIM temperatures.

The limited resolution and effective area of the current generation of X-ray grating 
spectrometers makes the detection of X-ray absorption lines challenging.
To date, there have been only a handful
of reported detections of X--ray lines from the WHIM, typically from \ovii\ and \oviii.
These detections include absorption features toward the targets
 H~2356-309 \citep{buote2009,fang2010}, PKS~2155-304 \citep{fang2002,fang2007,yao2009},
Mkn~421 \citep{nicastro2005,rasmussen2007,yao2012}, Mkn~501 \citep{ren2014} and
1ES~1553+113 \citep{nicastro2013}. Given the limited statistical significance of
all X--ray lines detected to date, it is important to investigate additional
sightlines and understand the correlation between UV and X--ray absorption lines.

In this paper we investigate the presence of X--ray absorption lines from the WHIM
towards the source \pg, a quasar at $z=0.177$. This sightline is 
 well studied in the FUV \citep[e.g.,][]{sembach2004,lehner2007,danforth2010,tilton2012}
and it is
rendered particularly
interesting by the presence of several \hi\ broad Lyman--$\alpha$ absorption (BLA) lines
and \ovi\ absorption line systems at $z>0$ reported by \cite{tilton2012} from 
Hubble Space Telescope (\hst) STIS and COS data. In particular,
the BLA at $z=0.0928$ detected from \hst\ data is one of the broadest \hi\ lines
in the \cite{tilton2012} sample ($b=130$ km/s), and it is indicative
of gas at $\log T(K) \simeq 6$ if the line width is purely thermal. 


This paper is structured as follows. In Section~\ref{sec:pg} we describe
the source \pg\ and the
available FUV data, in Sections~\ref{sec:X-ray} and \ref{sec:results} we describe the analsis of the \chandra\
and \xmm\ data, in Section~\ref{sec:interpretation} we provide
our interpretation of the X--ray and FUV results, and in Section~\ref{sec:conclusions}
we present our conclusions.
Throughout this paper we assume a standard $\Lambda$CDM cosmology with $h=0.7$ and $\Omega_{\Lambda}=0.7$.

\begin{figure}[!ht]
\centering
\includegraphics[width=5in,angle=-0]{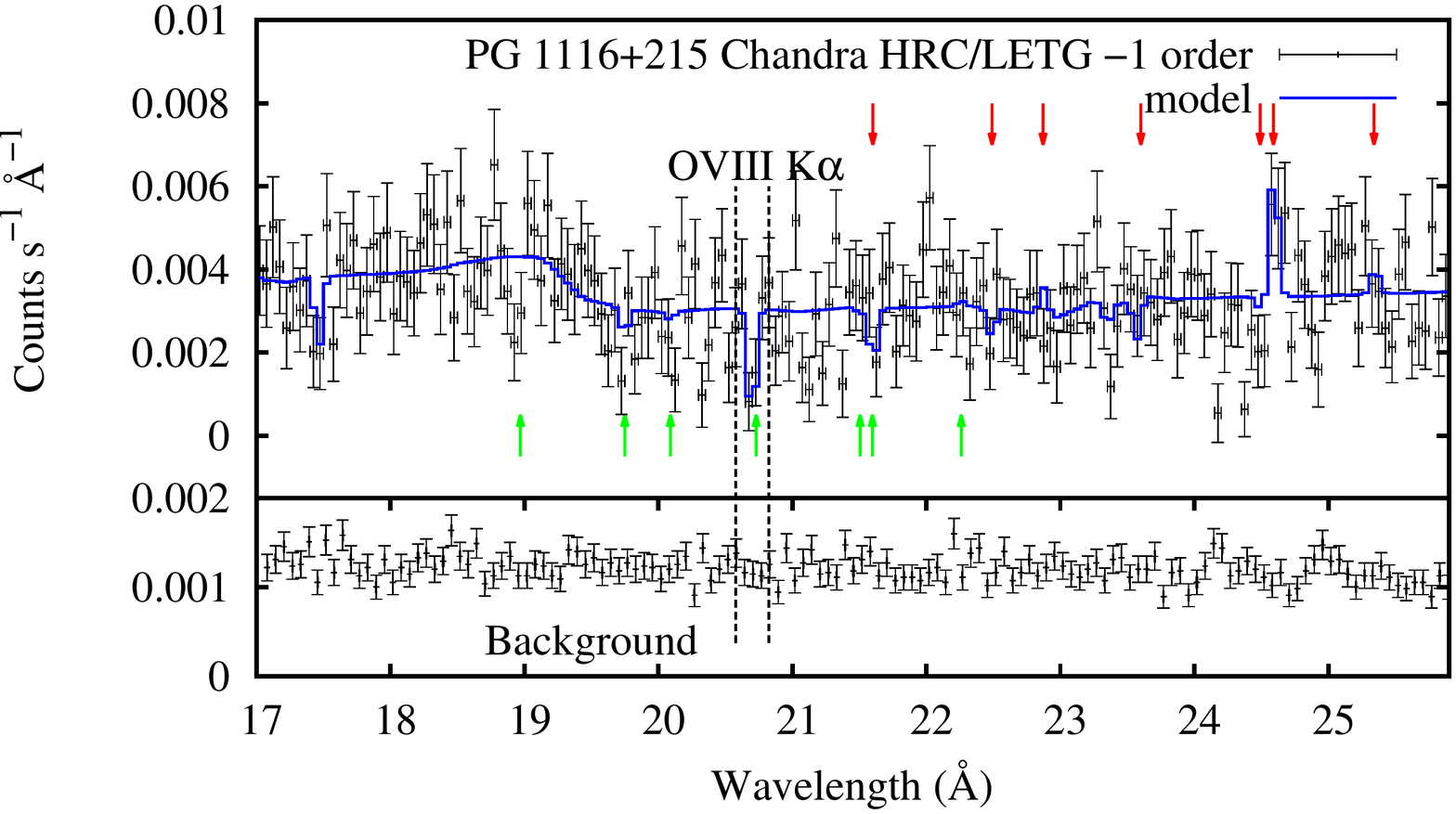}
\vspace{-1cm}
\includegraphics[width=5in,angle=-0]{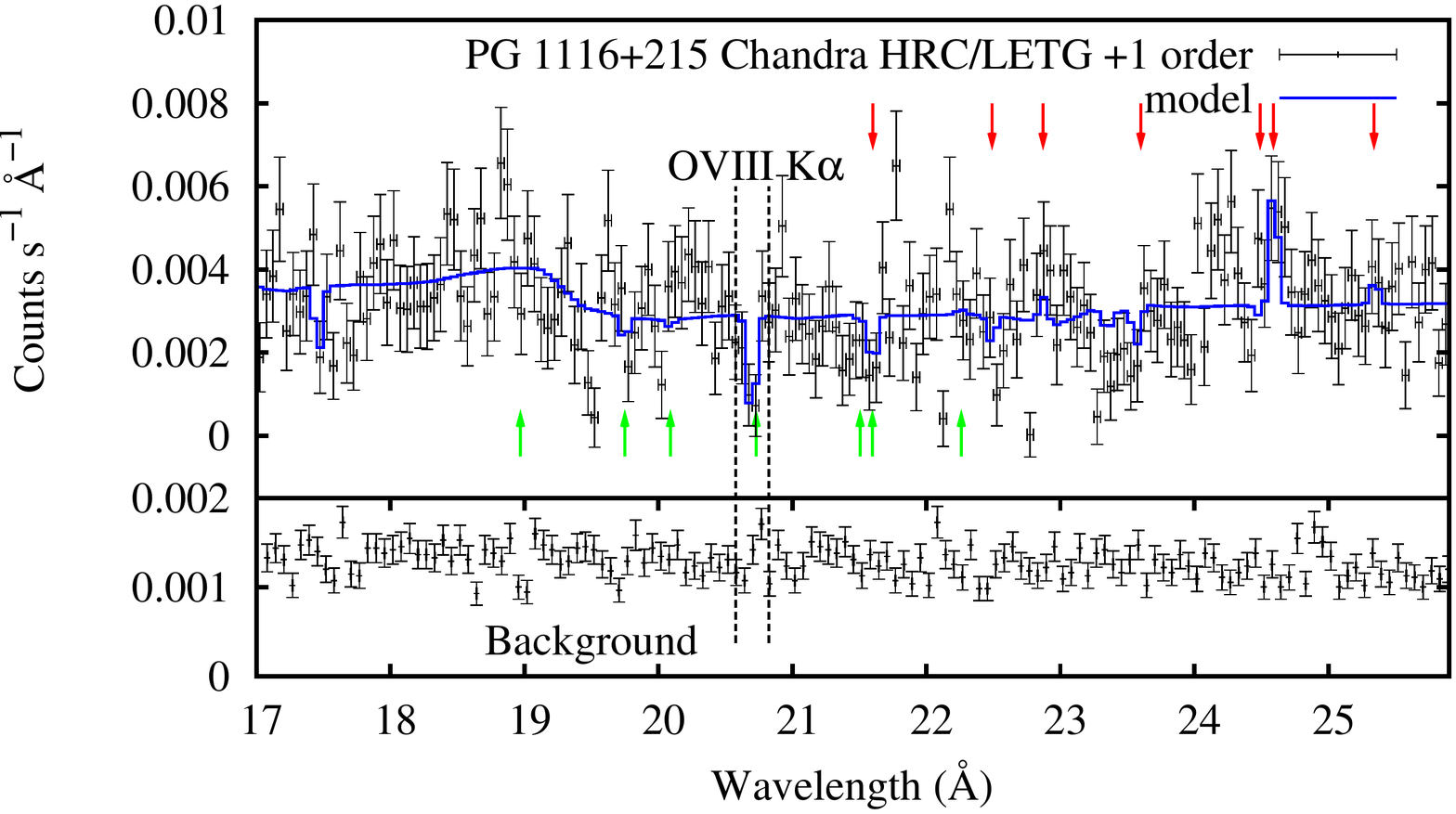}
\label{fig:letg}
\caption{LETG spectra of \pg. The spectra were binned to a 0.05 \AA\ resolution,
and the error bars correspond to the Poisson errors accoding to
the \cite{gehrels1986} approximation. Several datapoints in both spectra 
correspond to a small number of counts ($\leq 10$), where the Gaussian
approximation of the Poisson distribution would  not be accurate.
Arrows mark the expected positions of the two \ovii\ and \oviii\ K$\alpha$
lines respectively at redshift 0, 0.041,0.059,0.0928,0.1337 and 0.1385.
The background (bottom panels) is extracted from a region 
10 times larger than that of the source and it accounts on average for $\sim 30$~\% of the
total count rate. The feature at $\lambda \simeq 17.5$ is a possible \oviii\ K$\beta$
discussed in Section~\ref{sec:Kbeta}.} 
\end{figure}

\begin{figure}[!ht]
\centering
\includegraphics[width=4.5in,angle=-0]{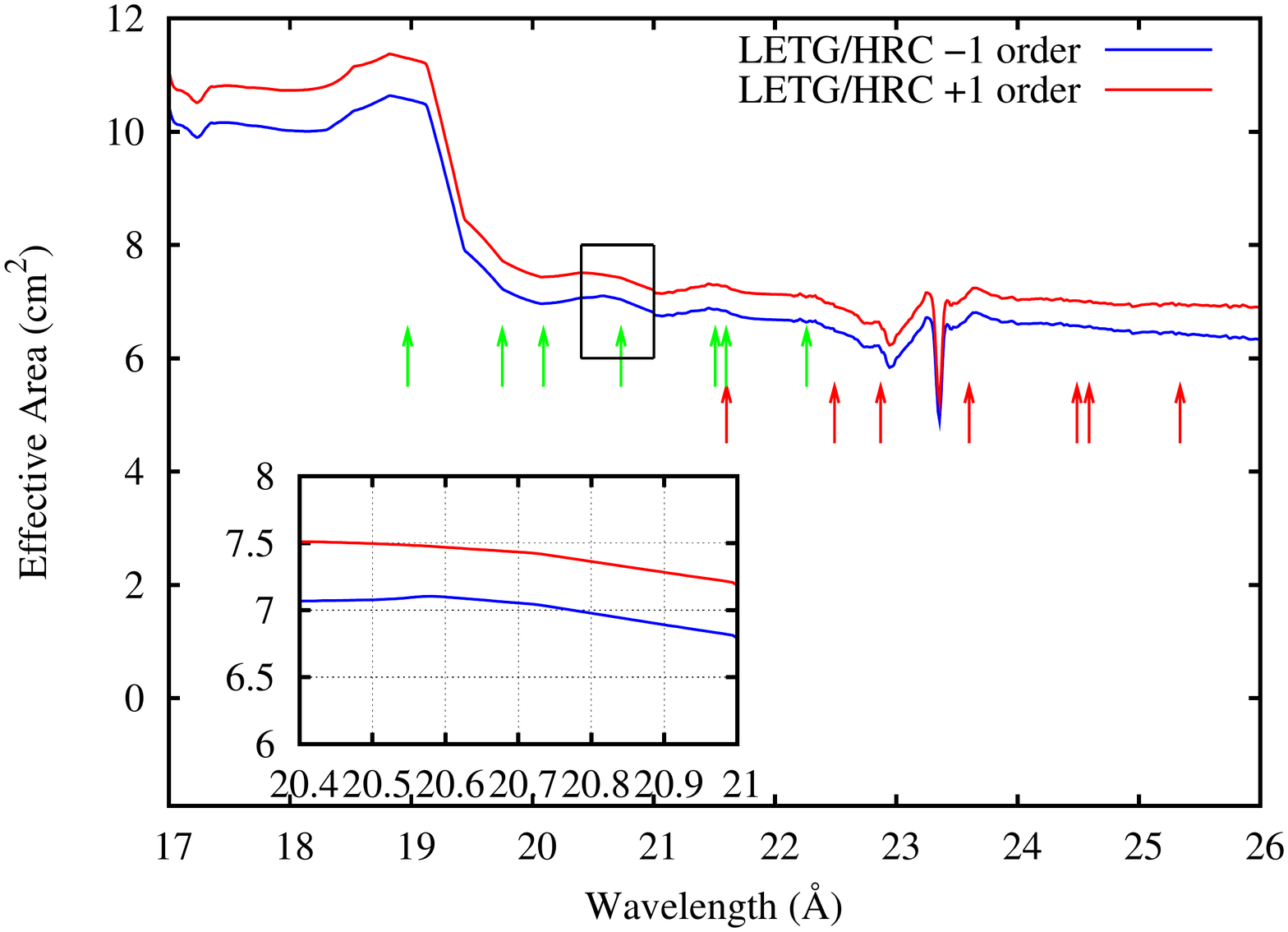}
\vspace{-1cm}
\includegraphics[width=4.5in,angle=-0]{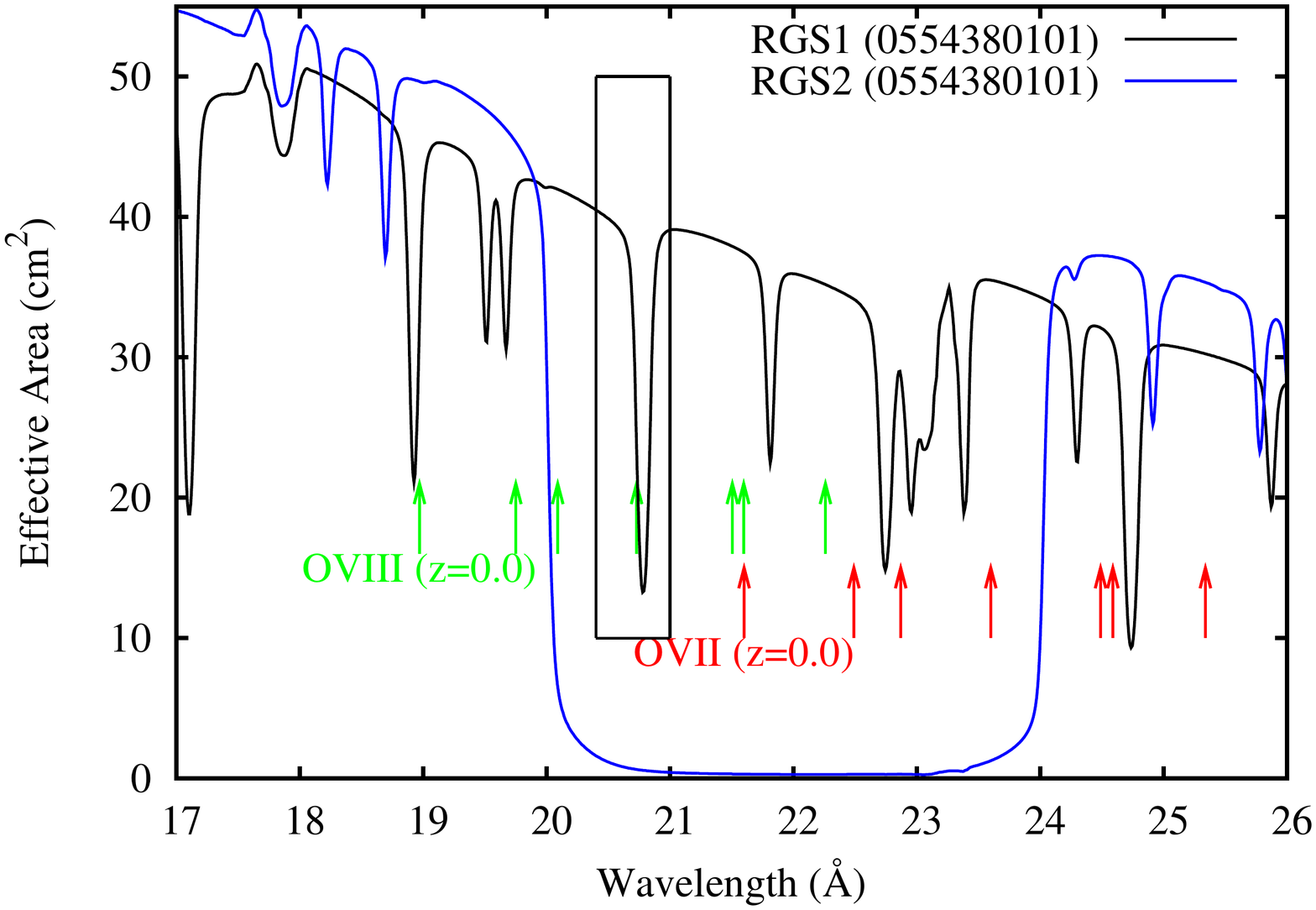}
\caption{Effective areas of the HRC/LETG instrument (top, $\pm1$ order shown separately) 
and of the RGS1 and RGS2 instruments (bottom, $\pm1$ order averaged for each instrument). 
The inset in the top panel is a zoom--in of the region of interest to the \oviii\
K$\alpha$ line at $z=0.0911$ (black box), showing that the HRC/LETG has uniform effective area there.
Several instrumental edges and 
the lack of effective area between 20-24 \AA\ for RGS2 limit
the use of the \xmm\ data for this study of \pg, including the \oviii\
K$\alpha$ line at $z=0.0911$ which falls in a region (black box) 
where the effective area drops by a factor of 4.
Arrows mark the expected positions of the two \ovii\ and \oviii\ K$\alpha$
lines respectively at redshift 0, 0.041,0.059,0.0928,0.1337 and 0.1385.}
\label{fig:effarea}
\end{figure}



\section{The sightline towards \pg}
\label{sec:pg}
\pg\ is a quasar at redshift $z=0.177$. \fuse\ and \hst\ FUV data were
analyzed by \cite{tilton2012}, who detected several absorption lines 
from intervening gas towards the source. Among the absorption lines detected in the FUV, 
we are especially interested in following up those absorption--line 
systems that have the potential
for associated X--ray lines, in particular
the two K$\alpha$ resonant lines from oxygen,
\ovii\ ($\lambda=21.602$ \AA)
and \oviii\ (doublet at a center wavelength of $\lambda=18.969$ \AA).
The atomic data relevant to this paper are presented in Table~\ref{tab:atomic}.
Oxygen is expected to be the most abundant 
element with atomic number $Z>2$ in the interstellar medium, and \ovii\ and \oviii\ are the most 
abundant oxygen ions in collisional ionization equilibrium 
at $\log T(K) \geq 6$ \citep[i.e., the high--temperature end of the WHIM range,][]{mazzotta1998}.

\begin{table}[!t]
\caption{Atomic data for relevant resonance lines, from  \cite{verner1996}.}
\label{tab:atomic}
\centering
\begin{tabular}{lllllll}
\hline
Ion & Line Transition &  Name	&Wavelength (\AA) & X--ray Energy (eV)& Osc. strength \\
\hline
H~I	& $1s-2p$ 	& Lyman-$\alpha$	& 1215.67$^{\star}$	& \nodata	& 0.416 \\	
O~VI	& $1s^2 2s - 1s^2 2p$ &  & 1037.6,1031.9 & \nodata &  0.199 \\
O~VII 	& $1s^2 - 1s2p$ &  K$\alpha$	& 21.602 	& 574.02  	& 0.696 \\
O~VIII  &  $1s-2p$       &  K$\alpha$     & 18.969$^{\star \star}$	&653.66	&  0.416 \\
O~VIII  &  $1s-3p$       &  K$\beta$     & 16.006$^{\star \star \star}$      &0.7747 &  0.079 \\
\hline
\end{tabular}

\flushleft $\star$ The \hi\ Lyman-$\alpha$ is a  doublet of $\lambda\lambda\; 1215.6736,1215.6682$.\\
\flushleft $\star\ \star$ The \oviii\ K$\alpha$ is a doublet of $\lambda\lambda\; 18.9725,18.9671$.
\flushleft $\star\ \star \star$ The \oviii\ K$\beta$ is a doublet of $\lambda\lambda\; 16.0067,16.0055$.
\end{table}

For this purpose, we select two classes of absorption--line systems towards  \pg\ among those
 detected by \cite{tilton2012}:
\ovi\ systems that have both lines in the doublet detected, and broad \hi\ Lyman-$\alpha$ absorption
lines (BLA) with $b \geq 80$ km/s. \ovi\ systems are traditional
signposts for the WHIM, since \ovi\ is the most abundant ion with
strong resonance lines in the FUV \citep[e.g.][]{shull2012}. 
It is clear that, for a single--temperature WHIM cloud in collisional ionization equilibrium (CIE),
\ovi\ and \ovii\ will coexist in significant amounts only in a very narrow range
of temperatures, and one does not expect virtually any \oviii\ at temperatures where \ovi\
is present. Nonetheless, it is possible that WHIM clouds have a multi--temperature structure,
as speculated by \cite{shull2003}, and therefore \ovi\ is a useful indicator
of WHIM at higher temperatures.
BLA's are also potential indicators of hot gas,
with a thermal broadening of $b \geq 80$ km/s indicating a temperature of $\log T(K) \geq 5.6$.
In this temperature range, 
we may in fact expect to find significant column densities of the  \ovii\ and \oviii\ ions.
The FUV absorption lines that meet these criteria are listed in Table~\ref{tab:tilton}.

In particular, we note that the $z=0.0928$ BLA has the highest Doppler $b$ parameter
of the entire sample studied by \cite{tilton2012}. 
For this paper we have re-analyzed the STIS and COS data of 
that absorption feature using the latest version of the pipeline used 
in \cite{tilton2012}, which implements minor improvements in exposure 
coaddition and continuum placement.~\footnote{See 
 http://casa.colorado.edu/~danforth/science/cos/costools.html for COS coaddition code. We
 used version 3.3 of \emph{coadd\_x1d}.}
  Fitting a single Voigt component to the COS data, we obtain
$z=0.0926814\pm0.000067$ (or $\pm$ 20 km/s), 
$b=153\pm17$ km/s and  $\log N_{HI} (\text{cm}^{-2})=13.26\pm0.06$.
Fitting a single Voigt component to the STIS/E140M data we measure
$z=0.0927801\pm0.000040$ (or $\pm$ 12 km/s), $b=126\pm14$ km/s and  $\log N_{HI}=13.43\pm0.05$.
 \cite{tilton2012}
reported  z=0.09279 ($\pm$ 15 km/s, or $z=0.09279\pm0.00005$), $b=133\pm17$ km/s and
 an equivalent width of $W_{\lambda}=0.11$ \AA,
which corresponds to an \hi\ column density of $N_{HI}=2.0 \pm 0.3\times 10^{13}$~cm$^{-2}$
(or $\log N = 13.30\pm0.06$). If the Doppler $b$ parameter is purely thermal,
a value of $b=133\pm17$ km/s corresponds to a temperature of $T = 1.06 \pm 0.27 \times 10^6$~K
(or $\log T (K) = 6.02\pm0.11$); for the single Voigt component fit to the COS data,
the temperature corresponding to the Doppler $b$ parameter is
 $T = 1.41 \pm 0.24 \times 10^6$~K
(or $\log T (K) = 6.15\pm0.10$).
It is worth noting that this BLA may have minor contamination 
from a weak foreground Galactic absorption line from C I $\lambda$1329. 
Though the COS data contain no suggestions of an additional, 
narrower absorption component, the higher-resolution STIS data contain a 
narrow feature of low significance, possibly corresponding to a C I component. 
Simultaneously fitting this feature with an additional Voigt component alters 
the BLA fit slightly to z=0.0927143 ($\pm$ 15 km/s), $b=139\pm17$ km/s, and $\log N=13.38\pm0.05$. 
The potential impact of C I contamination therefore appears 
to be negligible for the purposes of this study.
Because the re--analysis obtains results consistent with the original values from \cite{tilton2012},
we adopt those original values as listed in Table~\ref{tab:tilton}.

The X--ray spectra towards \pg\ have been previously used 
to search for absorption lines. \cite{fang2015}
studied $z=0$ absorption lines from warm--hot gas in the Galaxy but did not
investigate the $z>0$ systems that arise from the extragalactic WHIM.
 \cite{yao2009}   
focused  on selected \ovi\ systems detected with earlier FUV data,
including two systems at $z=0.059$ and $z=0.1358$ which we also investigate in this
paper, but did not analyze the $z=0.091$ system.
 \cite{yao2010} stacked \chandra\ data of several bright AGNs,
looked for lines at the redshift of bright nearby galaxies and likewise
did not investigate lines at $z=0.091$.
Neither paper reported any positive
identification of absorption lines from the extragalactic WHIM towards \pg. 
The prior knowledge on
the redshift of potential \ovii\ and \oviii\ systems that we have  from
the FUV data (Table~\ref{tab:tilton}) is essential
to search for faint X--ray absorption lines.
Given that we only seek to study lines at a known redshift,
the statistical significance of detection can be simply defined
as the ratio of the line flux $K$ and its 1-$\sigma$ uncertainty,
as discussed, for example, by \cite{nicastro2013}.
In the case of blind searches, i.e., without a prior knowledge on
the redshift, one needs to account for the number of independent detection opportunities
available for a given transition, leading to a reduction
in the significance of detection relative to the case of
lines with a redshift prior \citep{nicastro2013}.

\section{Spectral analysis of X--ray data}
\label{sec:X-ray}
In this paper we analyze the \chandra\ and \xmm\ X--ray grating spectra
of \pg. The \chandra\ HRC/LETG data provides a uniform coverage
at all wavelengths of possible \ovii\ and \oviii\ absorption lines of interest to this study. 
The \xmm\ data can be used to study only some of these
absorption lines, given a number of
instrumental features that make several wavelength intervals of interest
unavailable with the RGS spectrometer.

Spectral analysis was perfomed in XSPEC (version 12.8.2)  in the wavelength range 17--26~\AA\
where all relevant \ovii\ and \oviii\ K$\alpha$ lines from the systems listed
in Table~\ref{tab:tilton} are expected. For completeness,
we also investigate $z=0$ absorption lines from the same ions.
At the redshift of $z=0.1358$, the redshifted \oviii\ line position
coincides with the $z=0$ \ovii\ line. We therefore
set the flux of the redshifted $z=0.1358$ \oviii\ line to zero, and
let the flux of the $z=0$ \ovii\ line be free in the fit.
The continuum was
modelled with a power law in which both the normalization
and the spectral index was allowed to vary to find the best--fit parameters
and the significance of detection of the lines.
Each line was parameterized with
a Gaussian model that uses as parameter the total line flux ($K$, in units
of photons~s$^{-1}$~cm$^{-2}$),
redshift, line energy and line width
(parameter $\sigma_K$ of the Gaussian, in units of eV).

We use the Cash statistic $\mathcal{C}$ as the fit statistic, which is appropriate for
a data set of independent Poisson data points that may have bins with a low number of counts.
We prefer this over rebinning the spectra, in order to
retain a fixed bin width that matches the resolution of the \chandra\ data.
The $\mathcal{C}$ statistic is approximately distributed like a $\chi^2$ distribution
with $N-m$ degrees of freedom, where $N$ is the number of datapoints and $m$ is the number
of (interesting) fit parameters \citep{bonamente2013book}.
To determine the 1-$\sigma$ uncertainty in the flux $K$ of the lines and in
the other free parameters we therefore vary each interesting parameter until $\Delta \mathcal{C}=1$,
and use $\nicefrac{1}{2}$ of this range as the 1-$\sigma$ uncertainty, following
the same procedure that applies to Gaussian datasets with $\chi^2$ as the fit statistic.
In Section~\ref{sec:systematics} we further discuss the effects of fitting the \chandra\ data using
a variable bin size with a minimum of 25 counts per bin and
the $\chi^2$ fit statistic.

\subsection{Chandra}
\label{sec:chandra}

\chandra\ observed \pg\ for a total of 88~ks of clean exposure time with
the HRC/LETG spectrometer (observation ID 3145). 
Data reduction was performed in CIAO 4.7
using the standard processing pipeline (\textit{chandra\_repro}) 
that generates source spectra, background spectra and
response files. The +1 and -1 order spectra were kept separate
to better address the presence of spectral features in each
order spectrum.
The spectra were rebinned to a fixed bin size of 0.05~\AA,
 roughly the spectral resolution of the instrument.

The redshifts of the lines were fixed at the values of the expected lines
using the \hst\ redshifts shown in Table~\ref{tab:tilton}.
In the case of $\geq 3 \sigma$ features, including the
the $z=0.0928$ absorber, we allowed the redshift to be free 
in the final analysis to allow for small adjustments around the expected value.
In Section~\ref{sec:results} we discuss the best--fit values of the
redshifts for the detected features, and compare them with the \hst\ a priori redshifts.
Given the spectral resolution of LETG, we fixed the
line width parameter $\sigma_K$ of all lines to a fiducial 
value that corresponds to a width of 100 km/s, or $\sigma_K=0.2$ eV, which is
a characteristic value for the thermal broadening of lines from ions
in the temperature range $\log T(K) = 6-6.5$.
The line width is significantly smaller than the intrinsic resolution
of the LETG spectrometer, which is of order 1 eV for the wavelength
range of interest. Small changes from the nominal value
used of 0.2 eV have therefore negligible 
effect on the fit results. We address 
the source of systematic error associated with this assumption 
in our assessment of the significance of detection
of the lines in Section~\ref{sec:systematics}.

\begin{table}[!t]
\caption{FUV absorption lines from \cite{tilton2012} investigated in this paper.}
\label{tab:tilton}
\centering
\begin{tabular}{llll}
\hline
Redshift & Line ID & Doppler $b$ (km/s) & $W_{\lambda}$ (m\AA) \\
\hline
0.13373 & HI Lyman-$\alpha$ (BLA) & $81\pm6$ 	& $82\pm6$ \\
0.09279 & HI Lyman-$\alpha$ (BLA) & $133\pm17$ 	& $111\pm14$\\
0.04123 & HI Lyman-$\alpha$ (BLA) & $89\pm10$	& $73\pm9$ \\
\hline
0.17340 & OVI 1032,1038 & $47\pm7,24\pm13$ 	& $60\pm10,28\pm18$\\
0.13848 & OVI 1032,1038 & $24\pm8,41$		& $65\pm8,43\pm10$\\
0.05927 & OVI 1032,1038 & $10\pm6,17\pm12$	& $25\pm5,22\pm11$\\
\hline
\end{tabular}
\end{table}

The \chandra\ spectra and the best--fit models are shown in Figure~\ref{fig:letg}.
The error bars shown in the spectrum are based on the Geherels approximation
of the Poisson errors \citep{gehrels1986}.
The results of our fit are shown in Table~\ref{tab:fit}. The best--fit 
statistic is $\mathcal{C}_{min}=380.3$ for 346 degrees of freedom.
The flux of \pg\ at the wavelength of the line
is measured as $F_{\lambda} = 4\pm0.4 \times 10^{-4}$ photons~cm$^{-2}$~s$^{-1}$~\AA$^{-1}$\
in the \chandra\ observation.
The equivalent width $W_{\lambda}$ of the lines are obtained
from the relationship $W_{\lambda} F_{\lambda} = K$, where $K$ is the total
line flux (see Table~\ref{tab:fit}). In Table~\ref{tab:fit} we also report
constraints on the equivalent widths of the lines. When the data indicate
a positive normalization for the Gaussian line model, we report a 90\% upper limit
obtained as the equivalent width corresponding to a value of the line
flux $K \leq 1.3 \sigma_K$, where $\sigma_K$ is the uncertainty in $K$.

\begin{table}[!ht]
\centering
\caption{Results of the fit to the \chandra\ spectra. 
}
\label{tab:fit}
\begin{tabular}{lcccc}
\hline
Line ID & Redshift & Flux $K$ ($10^{-6}$ phot cm$^{-2}$~s$^{-1}$)& $\sigma$ (S/N) & $W_{\lambda}$ (m\AA) \\
\hline
\ovii\ K$\alpha$& 0.0 & $-14.1\pm7.3$ 	& -1.9 & $35.3\pm18.3$ \\
\ovii\ K$\alpha$ & 0.041 & -$7.6\pm7.2$ & -1.1 & $19.0\pm18.0$ \\
\ovii\ K$\alpha$ & 0.059 & $7.4\pm8.9$	& +0.8 & $\leq 28.9$  \\
\ovii\ K$\alpha$ & $0.0911\pm0.0004$ & $-11.1\pm7.4$ & -1.5 & $27.8\pm18.5$ \\
\ovii\ K$\alpha$ & 0.1337 & $-3.5\pm8.6$& -0.4 & $8.8\pm21.5$ \\
\ovii\ K$\alpha$ & 0.1385 & $38.7\pm11.1$ & +3.5& \nodata \\
\ovii\ K$\alpha$ & 0.1734 &  $7.8\pm10.1$ & 0.8	& $\leq 32.8$ \\
\hline
\oviii\ K$\alpha$ & 0.0 & $-8.0\pm4.6$ 	&  -1.7 & $20.0\pm11.5$\\
\oviii\ K$\alpha$ & 0.041 & $-5.9\pm6.8$& -0.9 & $14.8\pm17.0$ \\
\oviii\ K$\alpha$ & 0.059 & $-1.8\pm8.3$& -0.2 & $4.5\pm20.8$ \\
\oviii\ K$\alpha$ & $0.0911\pm0.0004^{\star}$ & $-30.7\pm5.9$& -5.2 & $76.8\pm14.8$ \\
\oviii\ K$\alpha$ & 0.1337 & $-1.7\pm8.3$ & -0.2 & $4.3\pm20.8$\\
\oviii\ K$\alpha$ & 0.1385 & 0.0	& \nodata&\nodata\\
\oviii\ K$\alpha$ & 0.1734 &$1.7\pm8.9$	& 0.2 & $\leq 28.9$ \\
\hline
\end{tabular}

\flushleft $^{\star}$ The redshifts of the two $z \simeq 0.091$ lines were linked in the fit.
Systematic errors in $z$ are discussed in Sec.~\ref{sec:systematics}.
\end{table}

\subsection{XMM--Newton}
\label{sec:xmm}

\xmm\ observed \pg\ in four separate visits (observations 0201940101, 0554380101, 0554380201 and
0554380301) for a total of 372~ks
of clean exposure. Reduction of the data was performed with the \emph{SAS} software,
using the \emph{rgsproc} pipeline and the calibration data current as of June 2015. 
The products of the reduction are $-1$ order averaged spectra for each observation
and the matching background spectra and response matrices. 

Each spectrum was rebinned to a bin size of 0.05~\AA, approximately matching the
RGS resolution in the wavelength range of interest (17-26 \AA). The bin size
is therefore the same
as for the \chandra\ spectra. We followed the same 
method of analysis of the spectra as for the \chandra\ data.
The only difference in the analysis is the fact that several spectral regions
corresponding to the lines of Table~\ref{tab:tilton} are unavailable with RGS.
Figure~\ref{fig:effarea} shows the \chandra\ HRC/LETG and the 
\xmm\ RGS effective areas. The main instrumental feature is
the lack of effective area of RGS2 between 20--24 \AA, due to the failure of
one of the CCD's in the focal plane of RGS2. RGS1 has also a number of instrumental
edges that prevent an accurate determination of those lines that follow near sharp
 gradients in the effective areas. In particular, 
the $z \simeq 0.091$ \oviii\ absorption line detected with \chandra\
near $\lambda=20.7$ \AA\ falls at the edge of a sharp intrumental feature
of RGS1, and therefore
we cannot reliably address the presence of such line using the RGS data.

Results of our RGS spectral analysis are summarized in Table~\ref{tab:rgs}, where 
we have used a dash 
in correspondence of those lines that cannot be probed because of
these instrumental features.
The four $\pm 1$ order RGS1 spectra were fit simultaneously 
to the same model used for the \chandra\ data, and so were the four  $\pm 1$ order RGS2 spectra;
normalizations of the continua was left free among the observations to allow for
flux variations among the different spectra.
The goodness of fit was measured using the Cash statistic as
$\mathcal{C}_{min}=687.5$ for 650 degrees of freedom for the RGS1 spectra, and
$\mathcal{C}_{min}=444.1$ for 376 degrees of freedom for the RGS2 spectra.

\begin{table}[!t]
\centering
\caption{Results of the fit to the \xmm\ spectra. Units are the same 
as in Table~\ref{tab:fit}. 
}
\label{tab:rgs}
\begin{tabular}{lccccccc}
\hline
Line ID & Redshift & \multicolumn{3}{c}{RGS1} & \multicolumn{3}{c}{RGS2} \\
        & & Flux $K$ & $\sigma$ (S/N) & $W_{\lambda}$ (m\AA) &  Flux $K$ & $\sigma$ (S/N) &  $W_{\lambda}$ (m\AA) \\
\hline
\ovii & 0.0 & $-3.3\pm2.2$  	& -1.5 & $18.9\pm12.6$ & \nodata & \nodata & \nodata \\
\ovii & 0.041 & -$1.0\pm2.5$    & -0.4 & $5.7\pm14.3$  & \nodata & \nodata & \nodata \\
\ovii & 0.059 & \nodata	   	& \nodata & \nodata & \nodata &  \nodata & \nodata \\
\ovii & 0.0911 & $-4.2\pm2.4$	& -1.8 & $24.0\pm13.7$ & \nodata & \nodata & \nodata \\
\ovii & 0.1337 & $-3.1\pm3.0$	& -1.0& $17.7\pm17.1$ 	& \nodata & \nodata \\
\ovii & 0.1385 & \nodata	& \nodata & \nodata 	& $-1.9\pm3.3$ & -0.6 & $10.9\pm 18.9$  \\
\ovii & 0.1734 &  $-1.7\pm3.2$  & -0.5 & $9.7\pm18.3$ &  $-0.8\pm3.2$ & -0.3 & $4.6\pm18.3$\\
\hline
\oviii & 0.0 & \nodata		& \nodata & \nodata	& $-0.3\pm2.0$ & -0.2 & $1.7\pm11.4$ \\
\oviii & 0.041 & \nodata	& \nodata & \nodata	& $-2.1\pm2.5$ & -0.8 & $12.0\pm14.3$ \\
\oviii & 0.059 & $-1.0\pm2.1$   &-0.5 & $5.7\pm12.0$ &\nodata	& \nodata & \nodata \\
\oviii & 0.0911 & \nodata  	& \nodata & \nodata& \nodata	& \nodata & \nodata \\
\oviii & 0.1337 & $0.1\pm2.7$  	& 0.0 & $\leq 15.6$  & \nodata	& \nodata & \nodata \\
\oviii & 0.1385 & 0.0           &\nodata & \nodata   & \nodata	& \nodata & \nodata\\
\oviii & 0.1734 & $-2.7\pm2.7$  & -1.0 & $15.4\pm15.4$	& \nodata & \nodata &\nodata \\
\hline
\end{tabular}
\end{table}

\section{Results of the spectral analysis}
\label{sec:results}
In this Section we provide a detailed description of the two 
features that were detected at $\geq 3 \sigma$ confidence in the fixed--redshift
search, namely the absorption line at $\lambda \simeq 20.7$~\AA\ 
(Sec.~\ref{sec:absorptionLine})
and a possible emission line at $\lambda \simeq 24.6$~\AA\ (Sec.~\ref{sec:emissionLine}). 
We also provide an analysis of sources of systematic uncertainty
in the detection of the absorption line at $\lambda \simeq 20.7$~\AA\ 
in Sec.\ref{sec:systematics}. 

\subsection{Absorption line at $\lambda \simeq 20.7$~\AA}
\label{sec:absorptionLine}
The strongest feature detected in the \chandra\ data at \emph{a priori} FUV wavelengths
is an absorption line at $\lambda \simeq 20.7$ \AA, which corresponds
to that of K$\alpha$ absorption from \oviii\ at $z \simeq 0.0911\pm0.0004$.
The nominal significance of detection for this line
is 5.2~$\sigma$, and the feature is clearly visible in each of the $\pm 1$
order \chandra\ grating spectra (Figure~\ref{fig:letg}).
We do not detect significant absorption lines consistent with 
\ovii\ or \oviii\ K$\alpha$ originating from other a priori FUV redshifts.
 As we discussed
in Section~\ref{sec:xmm}, this feature cannot be measured
accurately by \xmm. 

The \cite{tilton2012} BLA absorption line has a measured redshift
of $z=0.09279\pm0.00005$,
which is confirmed by the re--analysis of the \hst\ data presented
in Section~\ref{sec:pg}. In the initial analysis of the spectra we had fixed the redshift
of the \chandra\ absorption line at this value and obtained a line flux of
$K=-21.1 \pm 5.8 \times 10^{-6}$~phot~cm$^{-2}$~s$^{-1}$, for a
best--fit statistic of $\mathcal{C}_{min}=348.6$ ($\Delta \mathcal{C} = +8.9$ for one additional
d.o.f. relative to the fit with variable redshift).
In that case, the absorption line was detected at a confidence level corresponding to $3.7 \sigma$,
instead of $5.2 \sigma$ for a variable redshift. In the final analysis we preferred to
use a variable redshift, to account for possible sources of systematic errors in the determination
of the redshift and a possible physical off--set between the \hi\ and \oviii\ lines.
We also measured the redshift separately from the $+1$ and $-1$ order LETG spectra following
the same fitting method as for the combined observations.
The $+1$ order spectrum measures $z=0.0913\pm0.0006$ and the $-1$ order spectrum
measures $z=0.0912\pm0.006$, i.e., both spectra have a line shift that is
consistent with the value of $z=0.0911\pm0.0004$  measured from the joint analysis
of the two spectra.

A significant source of systematic error in the measurement of the \chandra\ redshift
is the accuracy of the wavelength scale.
The wavelength scale of the
LETG/HRC detector is known to have errors of up to 0.05 \AA, with an rms uncertainty
of order 0.01 \AA.~\footnote{See http://cxc.cfa.harvard.edu/cal/letg/Corrlam/}
This rms uncertainty corresponds to a redshift error of approximately $\sigma_z=0.0005$.
 Adopting this source of systematic error, our measurement of
 $z=0.0911\pm 0.0004 \pm 0.0005$ becomes
consistent with the
\hst\ measurement of $z=0.09279\pm0.00005$ at  the $3 \sigma$ level.
The agreement in the measurement of the line shift between the $\pm1$ order spectra
suggests however that the difference between the X--ray and FUV redshifts
may be real, rather than associated with an uncertainty of the wavelength scale of
the \chandra\ detector. Such difference may result from
a supra--thermal velocity structure in the X--ray absorber, 
as discussed in Section~\ref{sec:interpretation}, where
we provide our interpretation
of this absorption lines system and the temperature constraints obtained 
from the \chandra\ and \hst\ measurements.

Our choice to set the redshift of the putative \oviii\ K$\alpha$ as a free parameter in the final
analysis of the \chandra\ data warrants a discussion of the 
significance of detection of this line.
The nominal significance of detection, obtained as the ratio of the best--fit value
and the 1-$\sigma$ error bar and allowing a free redshift, is 5.2 $\sigma$. This absorption feature was 
detected at $\geq 3 \sigma$ significance using the fixed redshift from FUV data (see Section~\ref{sec:chandra});
had such fixed--redshift analysis not provided a significant detection of the feature, 
this absorption line would have gone undetected by our analysis.
The difference between the best--fit redshift of $z=0.0911\pm0.0004$ and the fixed FUV redshift
of $z=0.0928$ corresponds to a velocity difference of $\sim 510\pm120$~km/s. Peculiar velocities
of few $\time 100$~km/s are reasonable for large--scale unvirialized structures such as
WHIM filaments, i.e., they may be due to accretion or shocks. The use of a free redshift in the analysis can therefore
be viewed as a means for obtaining the best possible estimate of line parameters,
while retaining a resonable physical association with the fixed redshift used in the line search.

A more conservative approach towards determining the statistical significance of this line
is to account for the total redshift path length $\Delta z$ of the line search.
The redshift range of our search can be defined
as  fourteen redshift intervals
of size $\pm 3 \sigma$ around the \ovii\ and \oviii\ wavelengths of Table~\ref{tab:tilton},
where $\sigma$ is the statistical plus systematic error, estimated at $0.00075+0.0005$~\AA\
according to the values of Table~\ref{tab:fit} and the systematic errors in Section~\ref{sec:systematics}.
In so doing, we account for the prior redshift information from FUV data
while allowing adjustments to the X--ray redshift to account for reasonable  instrumental
or physical redshift differences. 
 We define the number of independent detection opportunities within that path length as
\begin{equation}
n = \frac{\lambda \Delta z}{\Delta \lambda}
\label{eq:trials}
\end{equation}
where $\lambda$ is the rest wavelength on the line of interest
 and $\Delta \lambda$ the resolution
of the spectrometer  \citep[e.g., the FWHM of the line--spread function,][]{nicastro2013}.
Using the atomic parameters in Table~\ref{tab:atomic}  and a resolution of 0.05~\AA,
there are approximately 20 independent opportunities in our search space for detecting
an \oviii\ K$\alpha$ absorption line and 23 opportunities
for \ovii\ K$\alpha$, i.e., 43 independent opportunities to
detect any one of the fourteen possible absorption lines of interest.
If we approximate
the nominal probability of exceeding  $\pm 5.2\sigma$ for a Gaussian distribution as
 $P \simeq 0.0000001$ (this value is an only an approximation, due
to uncertainties in the numerical integration of the tail of the Gaussian function),
the corrected null hypothesis probability becomes $P \simeq 0.000004$, which corresponds to
the probability of exceeding the mean of a Gaussian distribution
by $\pm 4.6\sigma$. The  detection of the putative \oviii\ K$\alpha$ line
therefore remains very significant even if we account somewhat conservatively
for the entire redshift path of the search.

\subsection{Emission line feature at $\lambda \simeq 24.6$~\AA}
\label{sec:emissionLine}
The only other spectral feature detected at a significance $\geq 3 \sigma$
is an emission line near $\lambda=24.6$~\AA\ in the \chandra\ data,
corresponding to an \ovii\ K$\alpha$ line 
at $z=0.1385$ (Table~\ref{tab:fit}). At that  redshift \cite{tilton2012} detected
both components of an \ovi\ absorption doublet (Table~\ref{tab:tilton}).
Using a free--redshift fit to the \chandra\ data for this line,
we obtain line parameters of $K=38.6\pm11.3$ and $z=0.1393\pm0.0008$, i.e.,
a less than 1-$\sigma$ redshift deviation from the \hst\ redshift. The free--redshift fit
for this line results in a 3.4$\sigma$ significance of detection, similar to that
in the fixed--redshift analysis of Table~\ref{tab:fit}.
The RGS2 detector is the only \xmm\ detector that can accurately probe this line,
and the RGS2 does not show a clear evidence of an emission line at that
redshift (see Table~\ref{tab:rgs}).  

To determine whether the flux difference 
between the \chandra\ and \xmm\ data for this feature is statistically significant, 
we re--measure the line fluxes using $\chi^2$ as the fit statistic with a minimum of
25 counts per bin, so that Gaussian statistics apply (the fluxes in Tables~\ref{tab:fit} and
\ref{tab:rgs} were obtained using Poisson statistics). We obtain a \chandra\
flux of $K_{\text{Chandra}}=3.66\pm1.06 \times 10^5$~phot~cm$^{-2}$~s$^{-1}$ and an 
\xmm\ flux of $K_{\text{XMM}}=0.07\pm0.33 \times 10^5$~phot~cm$^{-2}$~s$^{-1}$,
for a difference of $\Delta K =3.73\pm1.11 \times 10^5$~phot~cm$^{-2}$~s$^{-1}$.
This is a 3.3 $\sigma$ deviation from the null hypothesis
that the two fluxes are identical, for a null hypothesis probability of 
approximately 0.05\%. We conclude that there is a statistically significant difference
between the fluxes measured by \chandra\ and \xmm.

The  \chandra\ emission line may be a
transient feature, since the \xmm\ and \chandra\ observations were taken at different times.
Such putative transient emission line would have to be associated 
with a region of very small size, such that
its high density can produce the narrow emission line and have sufficiently low
\ovii\ column density as to be unable to produce absorption in the \pg\ spectrum;
these properties are not those of the WHIM, which is a diffuse and low--density medium. 
Moreover, the \chandra\ spectrometer is slitless and only sources with an angular
extent smaller than the PSF of the telescope (of order 1 arcsec) will
show a high--resolution spectrum. Therefore any emission line source of
a size larger than $\sim$1 arcsec would not have been detectable with LETG.
It is also possible that this feature is an emission line
at the redshift of the AGN, $z=0.177$. At this redshift, a wavelength of
$\lambda \simeq 24.6$~\AA\ corresponds to a rest wavelength of $\lambda_0=20.9$~\AA,
which in fact corresponds to a Lyman-$\beta$ resonance line from \nvii\ \citep[e.g.][]{verner1996}.
However, a stronger \nvii\ Lyman-$\alpha$  emission line ($\lambda_0=24.78$ \AA) would be expected at
$\lambda=29.2$ \AA, but the \chandra\ and  \xmm\ data do not show evidence of an emission
line at that wavelength.
Given the limited
significance of detection and the inconsistency between \chandra\ and \xmm,
additional data at this redshift are needed before providing a conclusive
statement as to the origin of this putative transient \ovii\ emission line at $z=0.1358$.

It is also possible that the \chandra\ emission line feature is simply due to a Poisson
fluctuation.
Our method of analysis in fact searches for \emph{absorption lines} in selected
wavelength ranges and we have no a priori reason to expect an \emph{emission line}, given
the physical parameters of the WHIM under investigation (low density and large size),
i.e., we cannot use a priori information on the redshift.
In the absence of such prior redshift knowledge, 
the probability of detecting an emission feature of a given significance due to Poisson 
fluctuations depends on the total redshift path length of the search.
Using Equation~\ref{eq:trials}, a fiducial wavelength of 20 \AA\ and a resolution of 0.05~\AA,
there are approximately 40 independent detection opportunities for an emission line in our search space.
Since there is a nominal probability of $P=0.0006$ to exceed $\pm 3.4 \sigma$ for a Gaussian distribution,
the corrected null--hypothesis probability becomes $P=0.024$, or 2.4\%, which corresponds to
the probability of exceeding the mean of a Gaussian distribution by $\pm 2 \sigma$. 
There is therefore a non--negligible probability of a chance fluctuation that yields such an emission
line feature in our search.

We conclude that this emission line feature in the \chandra\ data is either a
narrow transient emission line that is not associated with the WHIM, or it is due to
a Poisson fluctuation. We therefore do not further consider this spectral feature in this
paper.

\subsection{Systematics}
\label{sec:systematics}

The putative \oviii\ K$\alpha$ absorption line at $z \simeq 0.0911$
is the strongest spectral feature detected in our spectra.
In Section~\ref{sec:results} we have already addressed the systematics 
in the measurement of the redshift associated with the
determination of the \chandra\ wavelength scale.
To address the robustness of this detection, we further performed a set of tests on
the data and the method of analysis used to obtain this result. They consist of the following:
(1) re--analysis of
the data using binned data with a minimum of 25 counts, following
a more traditional method of analysis that uses $\chi^2$ as
the fit statistic \citep[e.g.][]{yao2010,fang2015};
(2) re--analysis using different background levels, to assess the dependence
of the significance of detection on the background;
(3) use of different values of the line width parameter $\sigma_K$
to study its effect of the line detection.

\subsubsection{Fit with spectra rebinned to $\geq$25 counts}
\label{sec:25counts}
For the analysis with a minimum of 25 counts per bin, we grouped the \chandra\ data
with the \textit{grppha} FTOOL software, and re--fitted the data in \textit{XSPEC}
using $\chi^2$ as the fit statistic. The best--fit statistic was
$\chi^2_{min}=244.4$ for 245 degrees of freedom ($\chi^2_{red}=1.00$), 
which corresponds to a null hypothesis probability of 42.7\%.
The best--fit parameters of the continuum in the 17-26 \AA\ band
are $\alpha=2.87\pm0.14$ (power--law index) and $norm=3.09\pm0.23\times 10^{-3}$
(normalization, in \textit{XSPEC} units), which are indistinguishable
from those obtained from the $\mathcal{C}$ fit statistic method
($\alpha=2.90\pm0.13$ and $norm=3.21\pm0.22\times 10^{-3}$).
For the significance of detection of the \oviii\ K$\alpha$ absorption line,
we find a redshift of $z=0.0898\pm0.0014$ and $K=-22.9\pm6.9 \times 10^{-6}$, for a significance 
of detection 3.3 $\sigma$.

\begin{figure}[!t]
\centering
\includegraphics[width=5in,angle=-0]{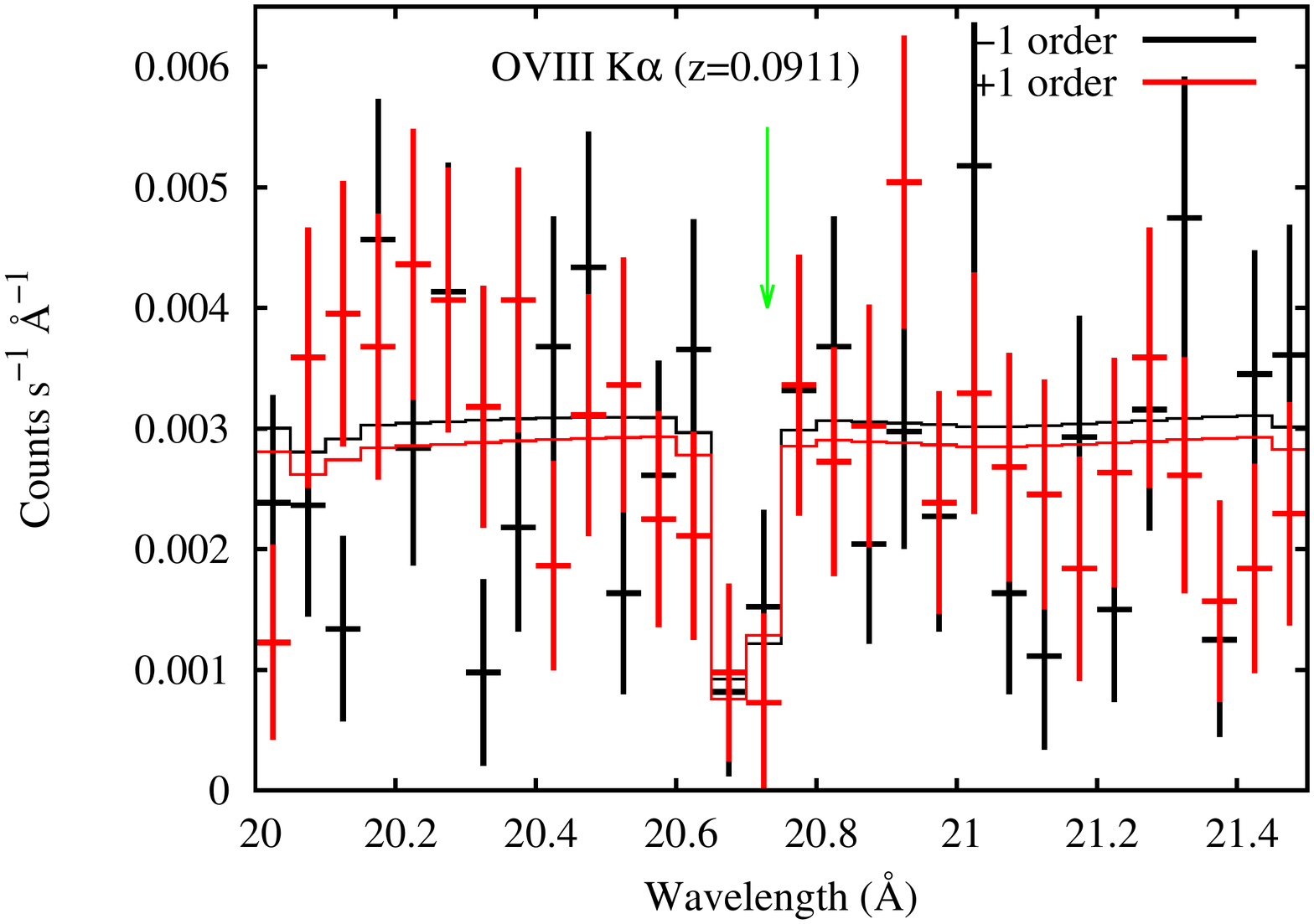}
\includegraphics[width=5in,angle=-0]{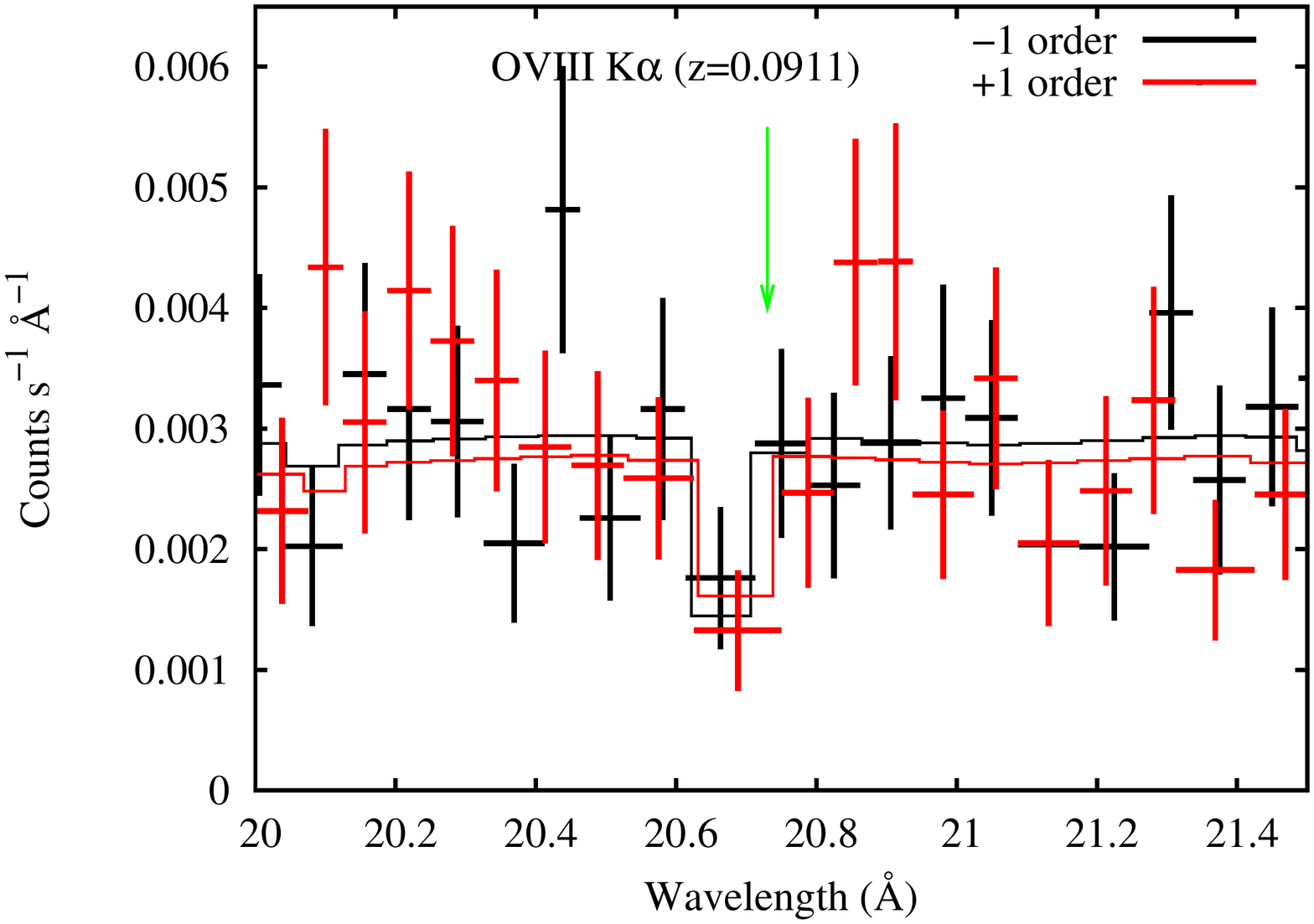}
\caption{Top: Zoom--in of Figure~\ref{fig:letg} in the region
of interest for the $z\simeq 0.091$ \oviii\ spectral region.
Bottom: Same spectral region as in the top panel, using 
the same \chandra\ spectra except for a re--binning to
have at least 25 total counts per bin (before background subtraction). 
Green arrows mark the position of the $z=0.0928$ \oviii\ K$\alpha$ line.}
\label{fig:zoom}
\end{figure}

The difference in the significance of detection between the two methods of analysis
can be explained with the differences in the spectral binning.
Each of the 0.05 \AA\ width bins
 contain significantly less than 25 counts, and therefore the two fitting methods 
($\mathcal{C}$ fit statistic with fixed bin width and $\chi^2$ statistic with
a minimum of 25 counts) are not equivalent. A comparison of the spectral bins used in the two
analysis is shown in Figure~\ref{fig:zoom}. The effect of rebinning to achieve a minimum of
25 counts is clear, with the bins being significantly wider than in the fixed--width Poisson fit method.
Some of the wider bins are an average of regions of continuum and regions where the
absorption line is present, and this limits both the ability to identify the line position (or redshift)
and its depth. Even with these limitations, the \oviii\ line is detected at $\geq 3 \sigma$ confidence level.
We conclude that the fixed bin--width method of Table~\ref{tab:fit} provides a more accurate method
of measuring the absorption line, while the 25--count bin method is still able to confirm the 
detection. The lower significance of detection with the latter method is clearly attributed to
limitations in the ability to measure the absorption line parameters due to the width
of the bins.

\subsubsection{Background analysis}
We address the effect of the background on the detection of the line by rescaling the
nominal background by $\pm 10$~\%, to account for possible systematics
of the HRC detector. We perform this test by
increasing and decreasing the background spectra
by 10\%, and re--fit the data to the same model as described in
the previous section. The effect of a background change by $\pm 10$~\% is negligible:
the best--fit $\mathcal{C}$ statistic is virtually unchanged
($\Delta \mathcal{C} \leq 2$ for both cases),
the best--fit redshift value remains the same as that in Table~\ref{tab:fit},
and the best--fit normalization becomes $K=-30.6\pm6.0 \times 10^{-6}$ (for -10\% background level)
and $K=-31.9 \pm 6.0  \times 10^{-6}$ (for +10\% background level), both negligible changes ($\leq 1$~\%)
relative to the nominal background. 

We also examine whether there are anomalous fluctuations in the background at the wavelengths of
the \oviii\ absorption line. In Figure~\ref{fig:letg} we show the background spectra for  our
\chandra\ LETG observations. The background is nearly constant across the wavelength range of interest,
and there are no statistically significant fluctuations from the average background level
at the wavelength of the \oviii\ absorption line
(marked by the fourth green arrow from the left and highlighted by the dashed lines).

We conclude that the background has a negligible effect on the significance of detection
of the \oviii\ absorption line.

\subsubsection{Effect of the line width parameter $\sigma_K$}
\label{sec:sigma}
The significance of detection of the line is likewise insensitive to the parameter
$\sigma_K$, given that the thermal
broadening of plasma at $\log T(K) = 6-6.5$ is significantly smaller
than the resolution of the \chandra\ LETG data. 
The width of the Gaussian line profile is proportional to the $b$ parameter via
the approximation of $\sigma_K \simeq (b/c) E$, where $E$ is the energy of the line
and $c$ is the speed of light. For the observed \oviii\ K$\alpha$ line, a value of
$b=100$ km/s corresponds to a line broadening of approximately $\sigma_K=0.2$ eV.

First we changed the fixed value of the width parameter to $\sigma_K=0.4$~eV
and  $\sigma_K=0.1$~eV (corresponding to a change of $\pm 100$~\% relative to the
nominal value of  $\sigma_K=0.2$), and re--fitted the spectra.
The best--fit statistic was unaffected ($\Delta \mathcal{C} \ll 1$ for both cases),
the best--fit redshifts and normalizations remained the same as those in Table~\ref{tab:fit}.
We also used a $\delta$-function model for each of the lines and repeated the fit.
The best--fit value of the parameters, their uncertainties and the
minimum fit statistic were again virtually identical to those obtained with
the nominal Gaussian models for the lines.

In an attempt to
determine observational constraints on the line width parameter from the \chandra\ data,
we also repeated the fits of Table~\ref{tab:fit}
with a Gaussian model of variable $\sigma_K$.
The best--fit model has a goodness of fit statistic $C_{min}=379.6$ for
345 degrees of freedom, for a decrease by $\Delta C_{min}<1$
using an additional free parameter. This reduction in the $C_{min}$ statistic
is not statistically significant.
We find a best--fit value of $\sigma_K=0.73\pm^{0.46}_{0.73}$~eV
and a normalization of $K=-32.5\pm7.7 \times 10^{-6}$, for a significance of
detection of $4.2 \sigma$, i.e., the K$\alpha$ lines remains statistically
significant even when the width of the Gaussian is a free parameter.
The range of allowed widths for the Gaussian line profile
corresponds  to  $b \simeq 360\pm^{260}_{360}$~km/s and it is therefore
consistent with a substantial amount of non--thermal broadening
(the thermal broadening of a 0.2 keV \oviii\ line is  50 km/s).

The fact that the fit statistic changes by $\Delta C_{min}<1$,
relative to the nominal value of $b=100$ km/s, for values 
in the range $b \simeq 0 - 520$~km/s indicates that
the \chandra\ data are insensitive to changes in the Doppler $b$ parameter 
and that the significance
of detection of the \oviii\ absorption line is unaffected
by changes to this parameter.

\section{Interpretation}
\label{sec:interpretation}
In this section we discuss possible interpretations for the
absorption line at $\lambda \simeq 20.7$ \AA, tentatively
identified as K$\alpha$ from \oviii\ at $z \simeq 0.091$.
First we constrain the physical parameters of the absorber
in Section~\ref{sec:physicalParameters}, then we analyze the presence
of the associated K$\beta$ line in Section~\ref{sec:Kbeta} and finally
we discuss the absorber's location in relation to known galaxy
structures in Section~\ref{sec:sdss}.

\subsection{Column density and temperature of the $z=0.091$ \oviii\ absorber}
\label{sec:physicalParameters}
The equivalent width of the \oviii\ K$\alpha$ line is
$W_{\lambda}=77 \pm 15$ m\AA\ (Table~\ref{tab:fit}).
For an optically thin absorber, the rest--frame equivalent width  $W_{\lambda}$ and
the column density $N$ (in ${\rm cm^{-2}}$) are related by
\begin{equation}
N = 1.13 \times 10^{20} \frac{W_{\lambda}}{\lambda^2 f} \text{ cm$^{-2}$}
\label{eq:N}
\end{equation}
where $f=0.416$ is the oscillator strength of the \oviii\ K$\alpha$ line,
$W_{\lambda}$ the equivalent width in \AA\
and $\lambda=18.97$~\AA\ is the rest--frame wavelength 
\citep[e.g.][]{verner1996}. Equation~\ref{eq:N} gives a column density 
of $N_{OVIII}=5.8\pm 1.1 \times 10^{16}$~cm$^{-2}$.

\begin{figure}
\includegraphics[width=4in,angle=0]{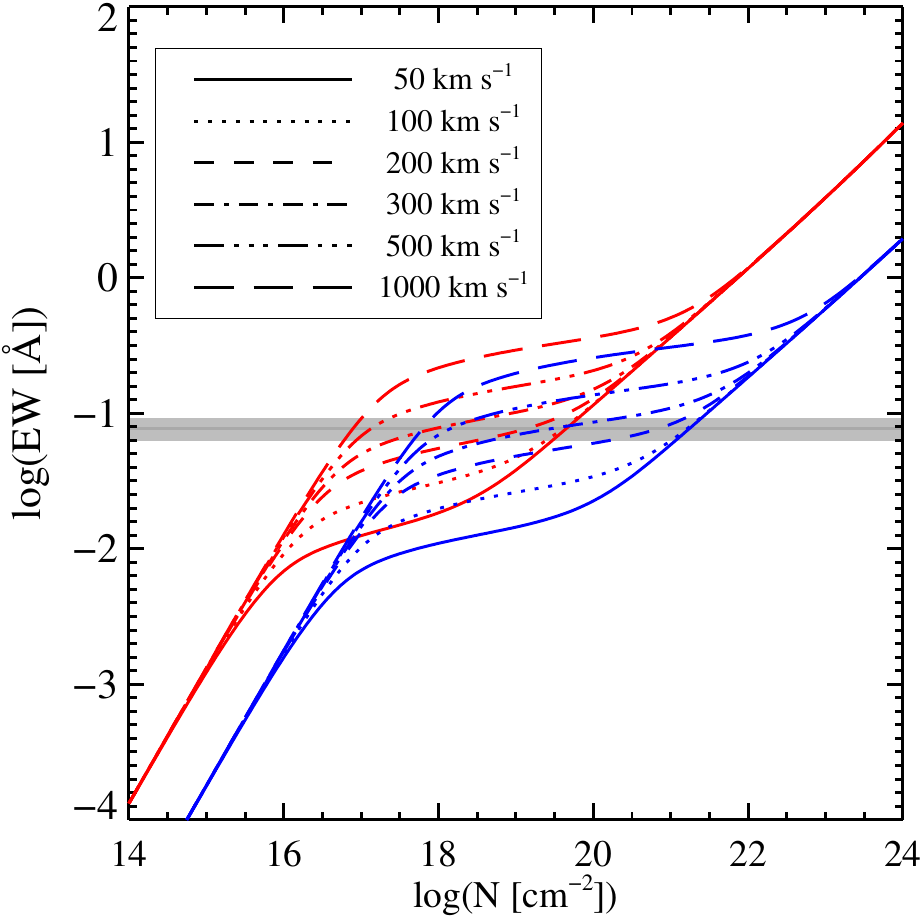}
\caption{Curves  of growth for te  \oviii\ K$\alpha$ (red) and K$\beta$ lines (blue). The grey
area marks the \chandra\ measurement of the \oviii\ K$\alpha$ absorption line.}
\label{fig:cog}
\end{figure}

Since the equivalent width of the line is larger than the 
thermal broadening (6 m\AA\ for $b=100$~km/s), we must examine the 
possibility of saturation of the line
 as function of the $b$ parameter.
The curves of growth for the \oviii\ K$\alpha$ line reported in Figure~\ref{fig:cog}
clearly show that, for a measured equivalent width of $W_{\lambda}=77\pm15$~m\AA,
an \oviii\ K$\alpha$ absorption line is optically thin only for  values $b \geq 500$~km/s
and it becomes increasingly saturated for lower values of $b$.
In Table~\ref{tab:cog} we report the measurement of the \oviii\ column density
as function of $b$ and the estimates for the equivalent width of the
associated \oviii\ K$\beta$ line, based on the curves of growth of Figure~\ref{fig:cog}.
Constraints on the \oviii\ K$\beta$ line are discussed in Section~\ref{sec:Kbeta}.
\begin{table}[!t]
\centering
\caption{\oviii\ K$\alpha$ column densites and predicted equivalent widths of
the associated K$\beta$ line using the curves of growth of Figure~\ref{fig:cog}.}
\label{tab:cog}
\begin{tabular}{lcc}
\hline
 b (km/s) & $\log N (cm^{-2})$ & K$\beta$ $W_{\lambda}$ (m\AA) \\
\hline
 50	& $19.65\pm^{0.16}_{0.19}$ 	& $18.3\pm1.5$ \\
 100 	& $19.61\pm^{0.17}_{0.23}$	& $30.2\pm1.6$ \\
 200	& $19.13\pm^{0.39}_{0.69}$ 	& $49.8\pm6.4$ \\
 300 	& $17.94\pm^{0.65}_{0.58}$ 	& $46.6\pm17.6$ \\
 500	& $17.13\pm^{0.24}_{0.20}$ 	& $20.2\pm8.8$ \\
 1000	& $16.89\pm^{0.11}_{0.12}$ 	& $13.3\pm3.4$ \\
\hline
\end{tabular}
\end{table}

In Section~\ref{sec:sigma} we have shown that the LETG data are largely insensitive
to the width of the absorption line and the data are consistent with a wide
range of $b$ parameters that includes substantial non--thermal broadening.
To estimate more accurately the line profile (including the $b$ parameter)
and column density of the \oviii\ K$\alpha$ line 
we make use of the \emph{SPEX} software package \citep{kaastra1996}, which has available
a collisional ionization equilibrium model that accounts for plasma temperature, ionization fractions, column densities,
line profiles, absorption edges and thermal/non--thermal broadening in a consistent way. 
For this purpose, we perform the following fits to the LETG spectra
 between 17-25 \AA, to include the
\oviii\ K$\alpha$ and K$\beta$ and the \ovii\ K$\alpha$ lines. 
Since the \oviii\ K$\alpha$ is the only line we
detect in our spectra, we consider a single--temperature CIE plasma in which oxygen 
is the only heavy element.
First we allow only thermal broadening of the lines and let the temperature of the
plasma and the redshift as free parameters. Then we repeat the same fits allowing for non--thermal
broadening of the lines (the $b_{NT}$ parameter, see Table~\ref{tab:spex}).
We used the combined $\pm1$ order spectrum, thus the fewer
degrees of freedom compared to the earlier fits.
We assumed a solar abundance of elements to convert the measured \oviii\ column density
to the total $N_H$. If a different abundance of oxygen relative to solar were used, 
e.g., if an abundance of $A=0.1$ solar, the $N_H$ column density would increase
by a corresponding amount, e.g., by a factor of 10 for $A=0.1$ solar.
The column density of \oviii\ is unaffected by the choice of metal abundance.

\begin{table}[!t]
\centering
\caption{Best--fit parameters of the 17-25 \AA\ \chandra\ spectra to
a CIE model in SPEX.}
\label{tab:spex}
\begin{tabular}{lccccc}
\hline
Fit     & kT (keV) 	& $b_{NT}$ (km/s)	& $N_H$ (cm$^{-2}$)	& $N_{OVIII}$ (cm$^{-2}$) & $C_{min}$ (d.o.f.) \\
\hline
1	& $0.82\pm^{\infty}_{0.43}$	& 0 (fixed)	& $4.8\pm^{\infty}_{4.5} \times 10^{22}$ & $5.0 \times 10^{17}$ & 168.2 (156) \\ 	 
2	& $0.34\pm^{0.56}_{0.10}$ 	& $647\pm^{370}_{386}$ & $6.1\pm^{133}_{4.1} \times 10^{20}$& $6.3\times 10^{16}$ & 165.4 (155) \\
\hline
\end{tabular}
\end{table}

An F--test for the significance of the additional component
in the fit with free $b$ relative to the one with fixed $b$ returns a value of $F=2.6$.
This value of the $F$ statistic has a null hypothesis probability
of approximately 90\%,
i.e., there is only a 10\% chance that the addition of the free parameter $b$ is not justified.
The SPEX fits therefore indicate that the data has a slight preference for an \oviii\ K$\alpha$ line
with a substantial amount of non--thermal broadening rather than a value of $b=100$ km/s;
this analysis is compatible with the 
phenomenological XSPEC fits  presented in Section~\ref{sec:sigma}.
This conclusion is borne out by the large value of $b$ and by the fact that, 
in the fit with thermal broadening alone,
the best--fit temperature is shifted to a larger value presumably to accommodate
a larger thermal broadening for the \oviii\ K$\alpha$ line. 
If the line is indeed broadened by $b\sim 500$~km/s, the line remains optically thin,
as indicated by the analysis of the curves of growth. In this case, the \oviii\
column density obtained by the SPEX fits (Table~\ref{tab:spex}) 
are consistent with the value obtained
from Eq.~\ref{eq:N}, i.e., the line is not saturated. The fit with a narrow line broadening results
in a very large total $N_H$ column density that cannot be easily explained using a WHIM filament
model, i.e., such large column density would be a typical column density associated with a cluster
(see Sec.~\ref{sec:sdss}). A significant broadening of the line is therefore a more plausible
scenario also because of its lower $N_H$ column density. 

The data also place a lower limit to the temperature, in particular
due to the available data
at the wavelengths of the \ovii\ K$\alpha$ line. 
 Using a 90\% confidence interval,
the temperature is constrained to $kT \geq 0.2$~keV, or $log T(K) \geq 6.4$,
from the fit with variable broadening parameter.
At this temperature, the thermal broadening of \oviii\ lines
corresponds to $b=50$~km/s, therefore non--thermal broadening
dominates if the temperature of the absorber is near this lower limit.
Temperature constraints
are independent of the overall abundance of oxygen relative to solar.
As shown in Table~\ref{tab:fit} and Table~\ref{tab:rgs}, 
\chandra\ and \xmm\ have weak absorption features
at the wavelength of the redshifted \ovii\ K$\alpha$ (respectively
1.5 $\sigma$ and 1.8 $\sigma$ significance). Deeper observations
may result in statistically significant detections of these lines
that would be very effective in constraining more accurately
the temperature of this absorption line system.



The Doppler parameter $b$  of the \hi\ BLA corresponds to
a lower temperature  than this value, 
since the largest estimate of $b$ corresponds to $\log T(K) = 6.15\pm 0.10$, 
as discussed in Section~\ref{sec:pg}. Therefore
the \hi\ BLA and the \oviii\ K$\alpha$ absorption lines
do not originate from an isothermal medium. 
In Section~\ref{sec:results} we have shown that the \hst\ and \chandra\
redshifts are within 3$\sigma$ of each other. 
Assuming that the centroid shift of the X--ray and FUV lines
are purely due to the Hubble expansion, the redshift difference indicates a 
$6\pm2$~Mpc line--of--sight separation between the two absorbers.
We examine this scenario in Section~\ref{sec:sdss}.


\subsection{Constraints on the \oviii\ K$\beta$ absorption line}
\label{sec:Kbeta}
The large values for the column density of the putative \oviii\ K$\alpha$ absorption line
suggest that the associated K$\beta$ line may be detectable with
the available \chandra\ and \xmm\ data. Both \chandra\ LETG and 
\xmm\ RGS1 have significant effective area at the wavelength of the redshifted \oviii\ K$\beta$ line
($\lambda \simeq 17.5$~\AA). In particular, the line falls in a wavelength range where RGS1
does not have any of the detector artifacts that prevented the investigation
of the K$\alpha$ line with this instrument.

\begin{figure}
\includegraphics[width=5in]{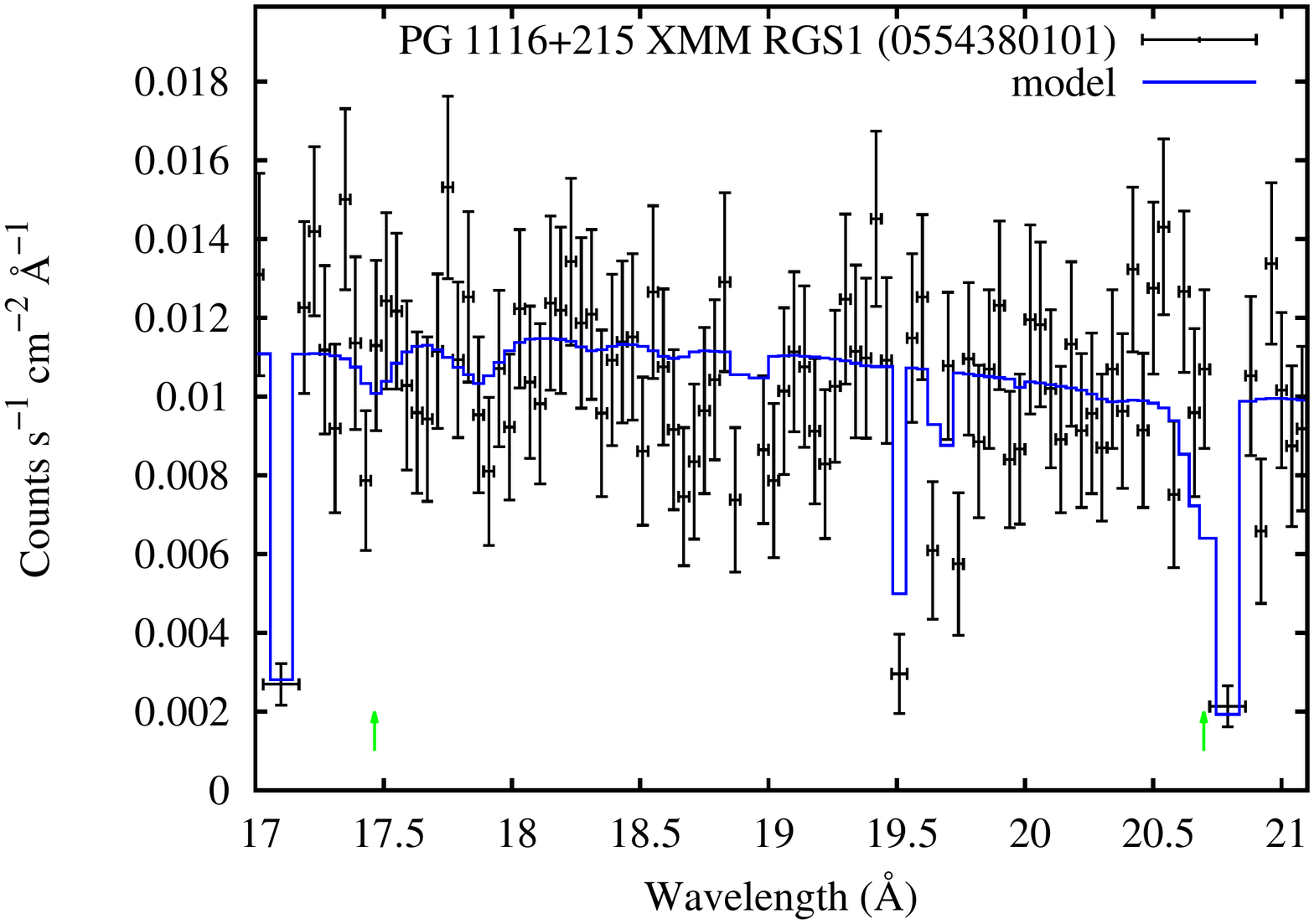}
\caption{RGS spectrum ($\pm1$ order, obs. 05543801010) fit to a power--law model
using the absorption model from the LETG fit of Table~\ref{tab:spex} with free
broadening parameter.  }
\label{fig:rgs}

\end{figure}

To determine the flux associated with \oviii\ K$\beta$ absorption at $z=0.0911$,
we repeated the same spectral fits 
with the addition of a K$\beta$ Gaussian absorption line model at the same redshift as
the K$\alpha$ line, with a variable normalization.
The results of these fits, the flux of the source at that wavelength
and the resulting equivalent width of the line are reported in Table~\ref{tab:Kbeta}.
For \chandra, the goodness of fit is $C_{min}=377.6$ for 345 degrees
of freedom, for a decrease  by $\Delta C_{min}=2.7$ using one additional
model parameter. For \xmm\ RGS1, the goodness of fit is $C_{min}=687.4$ for 649 degrees
of freedom, for a decrease  by $\Delta C_{min}=0.1$ using one additional
model parameter. For \xmm\ RGS2, the goodness of fit is $C_{min}=443.1$ for 375 degrees
of freedom, for a decrease  by $\Delta C_{min}=1.0$ using one additional
model parameter.
Neither instrument records a significant detection of the line, although the $\sim 2.4$~$\sigma$ 
significance of the normalization of the Gaussian model in the \chandra\ 
data is suggestive (see also Fig.~\ref{fig:letg}).
In Figure~\ref{fig:rgs} we show the RGS spectrum of one of the \xmm\ observations
using the same absorption model as the free-$b$ fit to the \chandra\ data of Table~\ref{tab:spex}.
The spectrum shows that the spectral region of the \oviii\ K$\alpha$ line is near a strong
detector absorption edge that prevented
an accurate measurement for this line and that the non--detection of the K$\beta$ line
is consistent with the model from the \chandra\ data.
The \chandra\ observation took place at a time when the \pg\ soft X--ray flux
was more than twice that in any of the \xmm\ observations. This is a possible reason
why the deeper \xmm\ observations were not able to provide a detection of the K$\beta$ line.

\begin{table}[!t]
\centering
\caption{XSPEC Fits to the  \oviii\ K$\beta$ absorption line at $z=0.0911$
and expected equivalent widths based on the K$\alpha$ measurement.}
\label{tab:Kbeta}
\begin{tabular}{lcccc}
\hline
	& K & $F_{\lambda}$ 	& \multicolumn{2}{c}{K$\beta$ $W_{\lambda}$ (m\AA)} \\
	& ($10^{-6}$ phot cm$^{-2}$~s$^{-1}$)  & ($10^{-6}$ phot cm$^{-2}$~s$^{-1} \AA$)			& Measurement & 99\% Upper Limit\\
\hline
LETG	& $-13.6\pm5.6$	& $4 \times 10^{-4}$ & $32.5\pm12.5$ &  $\leq$ 66.2\\
RGS1    & $0.0\pm2.5$  & $1.7  \times 10^{-4}$ &  $0.0 \pm 14.7$ & $\leq$ 36.9 \\
RGS2	& $-1.9\pm1.9$  & $1.7  \times 10^{-4}$ &  $11.2\pm11.2$ & $\leq$ 33.8\\
\hline
\end{tabular}
\end{table}

The curves of growth reported in Figure~\ref{fig:cog} were used to
predict the equivalent width for the K$\beta$ line
from the measured equivalent width for the K$\alpha$ line. Constraints on the
K$\beta$ equivalent widths from
LETG, RGS1 and RGS2 are shown in Table~\ref{tab:Kbeta}. The 99\% confidence
upper limits from our data 
are consistent with the 
equivalent widths expectations based on the K$\alpha$ measurements
reported in Table~\ref{tab:cog} for most values of the $b$ parameter.
In particular, the K$\beta$ constraints
are in agreement with the expectations for both
low values of the $b$ parameter ($b\leq 100$ km/s), and for large values
($b \geq 500$ km/s). The scenario of $b\leq 100$ km/s is problematic because of the
high column density implications, as discussed in Sec.~\ref{sec:physicalParameters}.
A value of $b \geq 500$ km/s explains the non--detection of K$\beta$ 
and implies a lower -- i.e., more reasonable  -- columnn density
for the  K$\alpha$ absorber.

We conclude that the current data do not have sufficient depth to achieve a significant
detection of the \oviii\  K$\beta$ line, regardless
of the degree of saturation of the K$\alpha$ line, and that the  K$\beta$ 
constraints suggest the presence of a substantial amount of non--thermal
broadening ($b \geq 500$ km/s) for the \oviii\ gas.

\subsection{Association with galaxies and filaments}
\label{sec:sdss}
\subsubsection{Galactic structures}

We  analyzed the available SDSS spectroscopy 
data in a region of $20 \times 20$ Mpc$^{2}$ around \pg.
The analysis consists in the identification and characterization
of possible
galaxy structures using the methods described 
in \cite{tempel2014a}.
The SDSS spectroscopic data at this redshift has an estimated completeness
of 20\% for galaxies at magnitudes $r<20$ and therefore our analysis
is hampered by such limitation.
The nearest galaxy to the sightline is more than 1 Mpc away from the sightline,
therefore a galactic origin for the FUV and X--ray absorption lines is unlikely.

\begin{figure}
\centering
\includegraphics[width=5in,angle=0]{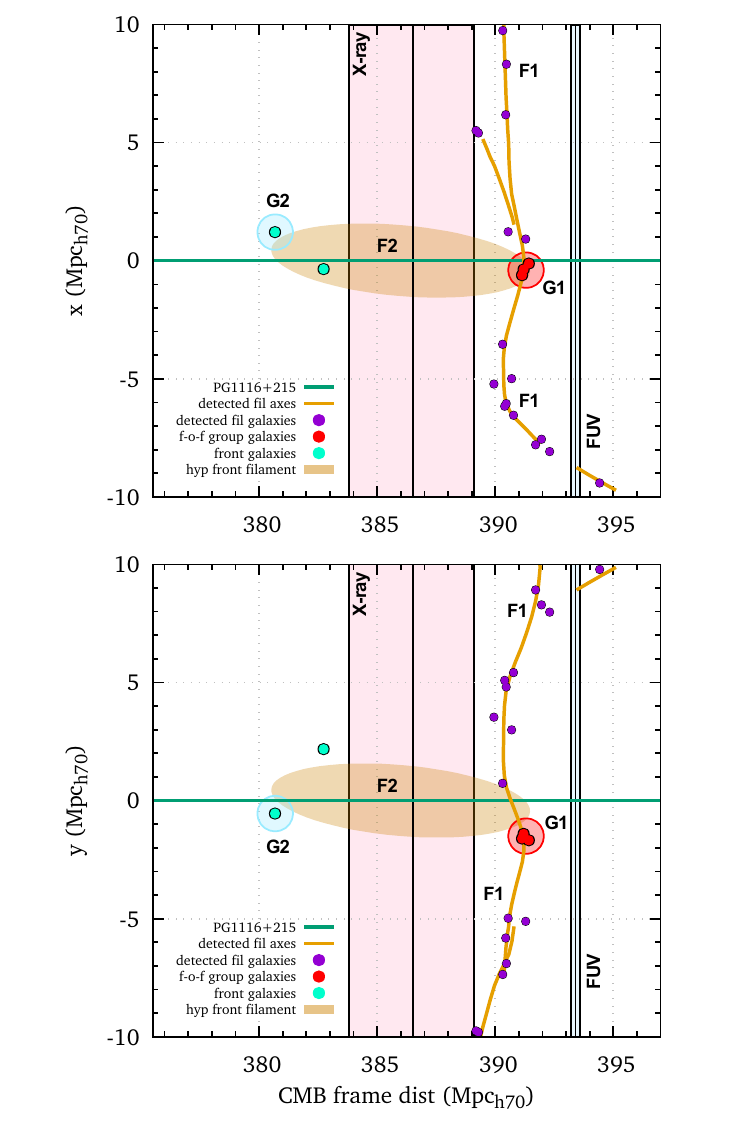}
\caption{Distribution of galaxies in the spectroscopic SDSS sample
along the \pg\ sightline around the implied X-ray and FUV absorbers
projected in two orthogonal planes. The distances corresponding to the
heliocentric redshifts of the X-ray and FUV absorption line centroids are
denoted with vertical lines. The significantly detected filament F1 is
denoted with a yellow line and purple dots. The friends-of-friends group G1
from \cite{tempel2014b}  and the compact Hickson group G2 \citep{mcconnachie2009}
are possibly connected by a filament (F2).}
\label{fig:sdss}
\end{figure}

The analysis of the large--scale SDSS data  reveals a
significant filament
of galaxies (F1, Fig.~\ref{fig:sdss}) that extends nearly perpendicular to the sight line
(i.e., the filament is primarily in the plane of the sky). 
Simulations of large--scale structure formation predict that filaments contain
most of the WHIM \citep{cui2012,nevalainen2015}.
This filament
intersects the sightline towards \pg\ and a three--galaxy
group (G1).
The redshifts of the galaxies that form this filament are between the redshift of
the FUV line and that of the \oviii\ K$\alpha$ line, indicating that
in principle both absorption lines may originate
from WHIM gas associated with this filament, provided that the absorbing gas moves with a
velocity of up to a few $\times 100$~km/s towards this filament, from either side
relative to our observing position. 
The current data do not permit an
accurate determination of the size of the filament in the direction of the sightline
towards \pg. The fact that the filament lies in the diretion \emph{perpendicular}
to this sight line would seem to indicate that its extent along the sight line
may be limited to a few Mpc.

We also find that three galaxies identified by \cite{tripp1998}
form a galaxy group, according to a friend--of--friends algorithm \citep{tempel2014b}.
The center of the group (G1, see Fig.~\ref{fig:sdss})
lies at a distance of $1.62$~Mpc from the
sightline towards \pg. Galaxy groups with a membership of a few galaxies
are not expected to have a soft X--ray halo that extends to such large distances.
In principle, this three--galaxy group could be part of a more massive cluster, if many
of its members have not been identified yet (e.g., due to the incompleteness
of the available SDSS data). To test this scenario, we analyzed the available
X--ray data at this location using publicly available ROSAT PSPC data
that had \pg\ at its aimpoint (25~ks of exposure, observation ID rp700228n00).
This galaxy group is too distant from \pg\ to be covered by \xmm\ and \chandra\
observations of the quasar.
At the location of this group, however, PSPC detects
no X--ray counts above the local background.
We conclude that it is unlikely
that any halo of hot gas associated with a small group at a distance of 2~Mpc
is responsible for the detected \oviii\ K$\alpha$ absorption.

The SDSS data further identify a galaxy at a projected distance of
$\sim$1.3~Mpc from the \pg\ sight--line and approximately 5~Mpc in front of the X--ray absorber.
This galaxy, according to \cite{mcconnachie2009}, is part
of a compact galaxy group (Hickson--type, or HCG)
with 4 galaxies identified through photometric redshifts (the
\citealt{mcconnachie2009} identifier for this group is SDSSCGB21597,
G2 in Fig.~\ref{fig:sdss}).
The presence of a galaxy group in front of the X--ray absorber
and another group nearly behind it suggests a scenario
of a possible WHIM filament (F2, see Fig.~\ref{fig:sdss}) 
extending between the two, with a distance
of 10 Mpc almost entirely along the sightline. 
This scenario would be the natural explanation for the broadening of the
\oviii\ K$\alpha$ absorption line reported in this paper. The Hubble flow
velocity difference at the ends of a 7~Mpc filament
is $\sim$500 km/s, consistent with the observations. 
In this scenario we have to assume a 
homogeneous distribution of WHIM over a pathlength of 7 Mpc.
This assumption, however, cannot be tested with the current data.

\subsubsection{Interpretation}

The $N_H$ values from Table~\ref{tab:spex}
can be used to estimate the density of the WHIM required to
create the observed absorption line.
Using the range of $L=1-10$ Mpc suggested by the SDSS data for filaments
along the sightline, the following relationship can be used to determine
the WHIM density,
\begin{equation}
\delta_b= \frac{N_H}{L} \left(\frac{\rho_{crit}}{m_H} \Omega_b\right)^{-1} 
= 100 \left(\frac{L}{\text{10 Mpc}} \right)^{-1} \left(\frac{N_H}{10^{21} \text{ cm$^{-2}$}} \right),
\label{eq:deltab}
\end{equation}
where we have used $m_H=1.67 \times 10^{-24}$~g, $\Omega_b=0.048$ \citep{Planck2015-cosmology},
$\rho_{crit}=6 \times 10^{-6}$~H atoms~cm$^{-3}$ and a somewhat conservative
 estimate of $N_H=10^{21}$~cm$^{-2}$. Equation~\ref{eq:deltab} therefore measures
 $\delta_b \simeq 100-1000$ for filament lengths in the range $L=1-10$~Mpc.
For filament lengths of $\sim 10$ Mpc (e.g., the possible filament F2), the temperature and 
density are therefore consistent with the high--density and high--temperature 
end of the WHIM that is
expected to provide \ovii\ and \oviii\ absorption lines. If the absorbing plasma
is more concentrated or resides in a filament with a 1--2 Mpc extent along the sight line
(e.g., filament F1),
the larger estimate of $\delta_b \sim 500-1000$ applies and the absorbing gas is consistent 
with a rare form of WHIM in the borderline 
between the intergalactic gas and the gas bound to galaxy clusters
\citep[see Fig. 6 of ][]{branchini2009}. Although the available X--ray data rule out
the presence of a massive cluster at this location,
a dense and hot WHIM in filament F1 is still a possible explanation
for the detected \oviii\ line, e.g., the high density may be related to
accretion of gas towards group G1.
Additional optical data in this area 
will overcome the incompleteness of the available SDSS coverage and improve our estimates 
of the geometry and extent along the sightline of the filament.
An estimate of the total column density of hydrogen associated with the \hi\ BLA detected
by \cite{tilton2012}
relies on an estimate of the temperature of the \hi\ absorber. If we assume a temperature
of $\log T(K)=6$, consistent with the Doppler $b$ parameter of the BLA, then
the \hi\ ionization fraction in CIE is $\log f_{HI}= -6.62$ \citep{gnat2007},
and the total column density associated with the BLA becomes $\log N_H = 19.92\pm0.06$.
This column is sufficiently lower than that associated with the \oviii\ K$\alpha$ line, so that
the estimates of Eq.~\ref{eq:deltab} are not significantly affected.

\section{Discussion and conclusions}
\label{sec:conclusions}
In this paper we have analyzed the \chandra\ and  \xmm\ X--ray grating spectra 
towards \pg, a quasar at $z=0.177$ with several intervening FUV absorption
lines detected in the \hst\ data \citep{tilton2012} that may be related to the WHIM.
We use the redshift of the FUV detections as priors for the search of \ovii\
and \oviii\ K$\alpha$  lines. 

The strongest feature we have detected
 is an absorption line at $\lambda \simeq$~20.7~\AA\ in the \chandra\ spectra
that we tentatively identify as a K$\alpha$ resonance line 
from \oviii\ at $z \simeq 0.091$.
This absorption line has a nominal statistical significance of $5.2 \sigma$,
and corresponds (within $3 \sigma$ of the \chandra\ redshift
measurements) to a very broad \hi\ Lyman-$\alpha$ (BLA) absorption line at $z=0.0928$
present in the \hst\ STIS and COS data. 
The \oviii\ K$\alpha$ detection
and the Doppler $b$ value of the \hi\ BLA indicate that
 the absorbing plasma is likely in a multi--temperature state. 
We conclude that we have a possible detection of a WHIM system towards \pg\
at $z\simeq 0.091-0.093$ that features \hi\ BLA (at $\log T(K) \leq 6.0\pm0.1$) 
and \oviii\ K$\alpha$ absorption
($\log T(K) \geq 6.4$). 
The temperature difference between the two phases is relatively small, if the Doppler
$b$ parameter of the BLA is an accurate indicator of its temperature. The energy required to heat
$\log T(K) = 6$ plasma to approximately $\log T(K) = 6.5$ may be provided by shocks
associated with the large--scale structure formation processes.

The available SDSS data provide evidence of galaxy structures towards \pg\
near the redshift of the \chandra\ and \hst\ absorption lines. In particular,
we have identified a large--scale filament that intersects the \pg\ sightline.
The putative WHIM associated with this filamente 
can in principle be associated with these absorption lines, assuming that the
absorbers are moving towards the filament with speeds of a few $\times $ 100 km/s.
We have also identified two galaxy groups that are located directly in front and 
behind the X--ray absorber. The available SDSS data are however not able
to determine whether there is a filament between them that may be responsible
for the \oviii\ absorption line. Deeper spectroscopic data to identify additional
galaxies in the field is necessary before we can make a more conclusive
association between the detected absorption lines and a specific galaxy structure.


The limited significance of detection of the putative \oviii\ absorption line 
at $z=0.0911$
reported in this paper requires 
a confirmation with additional data before its presence and the physical
state of the absorbing plasma can be determined accurately. 
Unfortunately,
the available \xmm\ data towards \pg\ cannot be used effectively to 
address the presence of this \oviii\ absorption line, since at
the wavelengths of this line RGS2 has no effective area, and RGS1 has
a significant drop in its efficiency due to an instrumental
feature. Additional observations will also address the presence
of \ovii\ K$\alpha$ lines. The available \chandra\ and \xmm\ data
in fact have absorption line features of respectively 1.5 $\sigma$ and 1.8 $\sigma$ 
at the wavelength of the $z=0.0911$ \ovii\ K$\alpha$ line that are
consistent with a significant column density of \ovii. These lines
may be detected with longer observations and potentially provide
the first X--ray system with multiple
absorption lines from the same element.

The detection of this putative \oviii\ K$\alpha$ absorption line in the \chandra\ 
spectra of \pg\ underscore
the importance of \hi\ BLA absorption lines as signposts of
the elusive high--temperature WHIM plasma \citep[e.g.][]{danforth2011,prause2007}. If this line
is confirmed, the X--ray data indicates the
presence of high--temperature WHIM with  a total hydrogen column density of
order $\log N_H \geq 20$.
 The combination of BLA's and X--ray data
for large samples has therefore the potential to identify large reservoirs
of warm--hot baryons, and possibly close the missing--baryons gap \citep{shull2012}.


Acknowledgements:
JN  is  funded  by  PUT246  grant  from Estonian  Research Council. 
We thank Dr. M. Weisskopf and Prof. A. Finoguenov for help in the interpretation of the data. 
TF was partially supported by the 
National Natural Science Foundation of China under grant No.~11273021.
ET and JT acknowledge the support by institutional research funding IUT26-2, 
IUT40-2 of the Estonian Ministry of Education and Research.
MB acknowledges support by the NASA/MSFC 2015 Faculty Fellow program.

\bibliographystyle{mn2e}
\bibliography{/home/max/proposals/max}

\end{document}